\documentclass[smallextended]{svjour3}       

\usepackage{amsmath}
\usepackage{amsfonts}
\usepackage{amssymb}

\usepackage{float}
\usepackage{graphicx}
\usepackage[percent]{overpic}
\usepackage[usenames,dvipsnames,svgnames,table]{xcolor}
\usepackage{booktabs}
\usepackage{tabularx}
\usepackage{url}

\usepackage{algorithm}
\usepackage{algpseudocode}

\begin{document}
\newcommand{\todo}[1]{\textcolor{red}{[TODO: #1]}}  
\newcommand{\annotate}[1]{\textcolor{ProcessBlue}{[COMMENT: #1]}}  
\renewcommand{\vec}[1]{\boldsymbol{#1}}  
\newcommand{\R}{\mathbb{R}}  
\newcommand{\E}{\mathbb{E}}  
\newcommand{\secref}[1]{Section \ref{#1}}  
\newcommand{\appref}[1]{Appendix \ref{#1}}  
\newcommand{\figref}[1]{Figure \ref{#1}}  
\newcommand{\tabref}[1]{Table \ref{#1}}  
\newcommand{\coderef}[1]{Listing \ref{#1}}  
\newcommand{\inline}[1]{\mintinline{python}{#1}}  
\newcommand{\specialcell}[2][c]{\begin{tabular}[#1]{@{}l@{}}#2\end{tabular}}  
\newcommand{\mc}[1]{\multicolumn{1}{c}{#1}}

\title{Synchronization Control of Oscillator Networks using Symbolic Regression}

\author{Julien Gout         \and
        Markus Quade \and
        Kamran Shafi \and
        Robert K. Niven \and
        Markus Abel
}

\authorrunning{Gout et. al.} 

\institute{J. Gout, M. Quade, M. Abel\at
              University of Potsdam \\
              Karl-Liebknecht-Str. 24/25 \\
              14476 Potsdam, Germany \quad and \at
              Ambrosys GmbH \\
              David-Gilly-Str. 1 \\
              14469 Potsdam, Germany \\
              \email{julien.gout@uni-potsdam.de}   \\
              \email{mquade@uni-potsdam.de}  \\
              \email{markus.abel@physik.uni-potsdam.de}
           \and
          K. Shafi, R. K. Niven \at
          School of Engineering and Information Technology \\
          University of New South Wales \\
          Canberra ACT 2600, Australia \\
          \email{k.shafi@adfa.edu.au} \\
          \email{r.niven@adfa.edu.au}
}

\date{Received: date / Accepted: date}

\maketitle
\begin{abstract}
Networks of coupled dynamical systems provide a powerful way to model systems with enormously complex dynamics, such as the human brain. Control of synchronization in such networked systems has far-reaching applications in many domains, including engineering and medicine. In this paper, we formulate the synchronization control in dynamical systems as an optimization problem and present a multi-objective genetic programming-based approach to infer optimal control functions that drive the system from a synchronized to a non-synchronized state and vice versa. The genetic programming-based controller allows learning optimal control functions in an interpretable symbolic form. The effectiveness of the proposed approach is demonstrated in controlling synchronization in coupled oscillator systems linked in networks of increasing order complexity, ranging from a simple coupled oscillator system to a hierarchical network of coupled oscillators. The results show that the proposed method can learn highly effective and interpretable control functions for such systems.

\end{abstract}

\section{Introduction}
The control of dynamical systems lies at the heart of modern engineering \cite{gautier-15}, and in many other disciplines, including physics \cite{PhysRevLett.64.1196,chen2003chaos} and medicine \cite{haken2006brain,schwalb2008history}.
This paper specifically focuses on the control of synchronization in dynamical systems.
Synchronization is a widespread phenomenon observed in many natural and engineered complex systems whereby locally interacting components of a complex system tend to coordinate and exhibit collective behavior \cite{Pikovsky2003Synchronization,strogatz2003sync}.
In dynamical systems, synchronization refers to the coordination phenomenon between multiple weakly coupled independent oscillating systems that influences the overall dynamics of the system. The role of synchronization control is to moderate this behavior (e.g., to drive the system into or out of synchronization) by applying an external force or control signal \cite{Pikovsky2003Synchronization}.
Synchronization control has significant implications for numerous application domains in engineering and science, including communications \cite{yang1997impulsive}, teleoperations \cite{li2011adaptive,shokri2014optimal} and brain modeling \cite{hammond2007pathological}, to name a few. A more specialized overview of synchronization in oscillators, and especially phase oscillators, can be found in \cite{dorfler2013synchronization}.

Several approaches exist for the control of dynamical systems, such as those based on control theory \cite{kirk2012optimal}, mathematical and numerical optimization \cite{nocedal2006numerical} and computational intelligence \cite{opt4ml} techniques. The ``optimal control" methods \cite{Becerra2008OptimalControl}, in particular, aim at driving and maintaining a dynamical system in a desired state. This is generally achieved by finding a control law, in the form of a set of differential equations, which optimizes (by maximizing or minimizing) a cost function related to the control task. For instance, in a medical application, the control of body tremors (e.g., due to seizures) may be achieved by minimizing the amplitude of body oscillations.

If the system is known in terms of a mathematical description, linear theory can be used \cite{Kirk1970OptimalControl,scholl2008handbook} in many cases to find the optimal control. However, for nonlinear, extended and consequently complex systems, linear theory may fail. In such cases, more general methods are needed to learn effective control laws. We have been looking for the most general method using analytical expressions, or in other words a method to infer control laws from an arbitrary domain which can be defined in a general way. We identified evolutionary machine learning methods as a suitable source of algorithms. Specifically, we describe the application of genetic programming (GP) \cite{Koza1992GeneticProgramming} to control synchronization in coupled networks, including a hierarchical network of coupled oscillators. Unlike neural networks and other black-box artificial intelligence methods that are commonly applied to optimal control, GP allows dynamically learning complex control laws in an interpretable symbolic form --- a method that is referred as symbolic regression \cite{schmidt2009distilling,vladislavleva2009order,Quade2016Prediction}.
Previously, evolutionary algorithms have been successfully used to optimize parameters for model-based control of synchronization \cite{shokri2014optimal,shokri2014comparison}. In contrast, we optimize the full expression, and not only parameters.

Previous attempts to use symbolic regression for the control of dynamical systems were mostly restricted to experiments
\cite{gautier-15,parezanovic2016,Duriez2017} without multi-objectivity and without the optimization of constants involved in the equations. In contrast, here we use a multi-objective formulation of GP, which allows learning much sparser, as well as multiple Pareto (or non-dominated) solutions \cite{schmidt2009distilling,vladislavleva2009order,Quade2016Prediction}. The Pareto solutions can be further analyzed for insight using the conventional analytical methods, such as bifurcation analysis --- subject of our ongoing work.
We demonstrate the effectiveness of the proposed control approach through application to different dynamical systems with growing level of complexity, ranging from a single oscillator to a hierarchical network. For each system, we show the useful optimal control terms found by symbolic regression to synchronize or de-synchronize the oscillators. The application to the control of synchronization in a hierarchical network of oscillators is motivated by brain disorder problems in the medical domain. Body tremors occur when firing neurons synchronize in regions of brain \cite{haken2006brain}. In a normal brain state, neurons are coupled to neighboring neurons such that adjacent neurons influence mutually. If the firing is periodic, which may appear due to the inherent dynamics of the excitable neurons, this mutual influence may give rise to synchronization \cite{Pikovsky2003Synchronization,Strogatz2006SyncBasin}. If the coupling term is very large, this synchronization may extend over a whole region in our brain and thus over many neurons. Eventually this collective firing leads to shaky movements of hands, arms or the head, and is treated as a brain disorder. One remedy to this problem is to implant a control device which resets the neurons and counteracts the collective synchronization. An evident question then is how to design such a controller which also minimizes design cost, energy consumption, or other medical constraints. We model this phenomenon, in this work, as a set of oscillatory units (representing neurons) coupled in a small hierarchical network, and apply the proposed approach for learning the optimal control laws for this model.

To the best of our knowledge, this is the is first application of a multi-objective GP to synchronization control in networks of coupled dynamical systems. The rest of the paper is organized as follows: \secref{sec:methods} provides the background information on control of dynamical systems, the common approaches for optimal control of dynamical systems, GP, and the specific methods used in our implementation of GP. \secref{sec:gpsetup} provides the details of the common GP parameters and other experimental settings used to evaluate the proposed method. The specific parameter settings are provided in each corresponding section. As a proof of concept, \secref{sec:examples} demonstrates the application of GP-based control using two simple benchmark dynamical system examples: a harmonic oscillator and the Lorentz system. GP is used to learn control to bring these system into a chaotic state or back. \secref{sec:results} presents our study of GP application to networked dynamic systems. Four systems are tested in this section, including a simple coupled oscillator system; and three systems of oscillators coupled, respectively, in a one-dimensional ring network; a two-dimensional torus network; and a hierarchical network. The paper concludes in \secref{sec:conc}.

black box
\section{Methods}
\label{sec:methods}

\subsection{Optimal control of dynamical systems\label{sec:feedback-control}}
\label{sec:oc}

\begin{figure}[htpb]
    \centering
    \includegraphics[width=1.\textwidth]{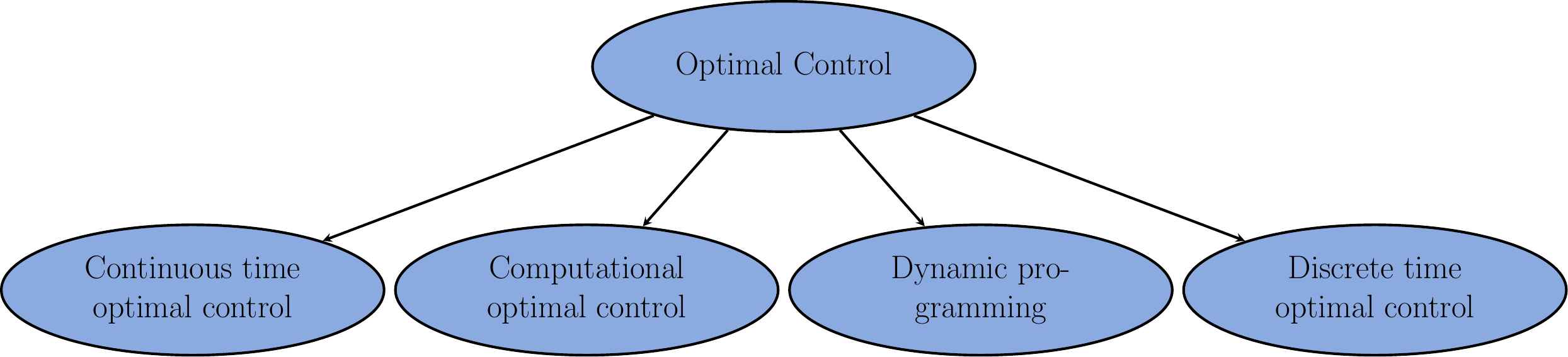}
    \caption{An overview of different types of optimal control. \label{fig:optimal-control}}
\end{figure}

The control of a dynamical system involves determining and manipulating the trajectory of the system in phase space in order to drive the system to a desired state. The control problem can be formulated as an optimization task with the objective to minimize a cost function defined in terms of the deviation of the state of the system from its desired one.
In general, a formal definition of an optimal control problem requires a mathematical model of the system, a cost function or performance index, a specification of boundary conditions on states, and additional constraints.

If there are no path constraints on the states or the control variables, and if the initial and final conditions are fixed, a fairly general continuous time optimal control problem reads: Find the control vector $\vec{u}:\R^{n_x} \times [t_s, t_f] \mapsto \R^{n_u}$ that minimizes the cost function
\begin{align}
    \Gamma = \varphi(\vec{x}(t_f)) + \int_{t_s}^{t_f} L(\vec{x}(t), \vec{u}(\vec{x}, t), t) \mathrm{d}t , \label{eq:oc-gamma}
\end{align}
subject to
\begin{align}
    \dot{\vec{x}} = \vec{\tilde{f}}(\vec{x}, \vec{u}, t),\,\, \vec{x}(t_s) = \vec{x}_s, \label{eq:oc-dynsys}
\end{align}
where $[t_s,t_f]$ is the time interval of interest;
$\vec{x} {:} [t_s, t_f] \mapsto \R^{n_x}$ is the state vector;
$\varphi : \R^{n_x} \mapsto \R$ is a terminal cost function;
$L { :} \R^{n_x} \times \R^{n_u} \times \R \mapsto \R$ is an intermediate cost function; and
$\vec{\tilde{f}} : \R^{n_x} \times \R^{n_u} \times \R \mapsto \R^{n_x}$ is a vector field.
Note that Eq.~\eqref{eq:oc-dynsys} represents the dynamics of the system and its initial state. This problem definition is known as the Bolza problem; and for $\varphi(x(t_f)) = 0$ and $\vec{u} = \dot{\vec{x}}(t)$ it is known as the Lagrange problem \cite{goldstine2012history}. Also note that the performance index $\Gamma$ is a functional, which is used to assign a real value to each control function $\vec{u}$ in a class.

The solutions to many optimal control problems cannot be found by analytical means. Over the years, many computational methods have been developed to solve general optimal control problems. The choice of a method for addressing an optimal control problem may depend on a number of factors, including the types of cost functions, time domain, and constraints considered in the problem. \figref{fig:optimal-control} shows different methods used in the optimal control of dynamical systems. Among these methods, the direct methods work by discretizing the control problem and solving it using nonlinear programming approaches. Some methods involve the discretization of the differential equations by defining a grid of $N$ points covering the time interval $[t_s, t_f]$, $t_s = t_1 < t_2 < \ldots < t_N = t_f$, and solving these equations using, for instance, Euler, trapezoidal, or Runge--Kutta methods \cite{press2007numerical}. In this approach, the differential equations become equality constraints of the nonlinear programming problem. Other direct methods involve the approximation of control and states using basis functions, such as splines or Lagrange polynomials.

The continuous-time problems mentioned above have discrete time counterparts. These formulations are useful when the dynamics are discrete (for example, a multistage system), or when dealing with computer controlled systems. In discrete time, the dynamics can be expressed as a difference equation:
\begin{align*}
    \vec{x}(k+1) = \vec{f}(\vec{x}(k), \vec{u}(k), k),\,\, \vec{x}(N_0) = \vec{x}_s ,
\end{align*}
where $k$ is an integer index, $\vec{x}(k)$ is the state vector, $\vec{u}(k)$ is the control vector, and $\vec{f}$ is a vector function.
The objective is to find a control sequence ${\vec{u}(k)}$, $k = N_0,\ldots, N_f - 1$, to minimize a performance index of the form:
\begin{align*}
    \Gamma = \varphi(\vec{x}(N_f)) + \sum_{k=N_0}^{N_f-1} L(\vec{x}(k), \vec{u}(k), k).
\end{align*}
See, for example, \cite{Lewis12995OptimalControl,Bryson1975OptimalControl} for further details on discrete-time optimal control.

Dynamic programming is an alternative to the variational approach to optimal control. It was proposed by Bellman in the 1950s and is an extension of Hamilton--Jacobi theory. A number of books exist on these topics including \cite{Lewis12995OptimalControl,Kirk1970OptimalControl,Bryson1975OptimalControl}.
A general overview of the optimal control methods for dynamical systems can be found in \cite{Becerra2008OptimalControl}. For further details, readers are referred to \cite{Athans2006OptimalControl}.

Unless otherwise stated, we approach the control problems presented in this paper using discrete-time numerical methods. However, for most purposes, the continuous-time formulation given in Eqs. \eqref{eq:oc-gamma} and \eqref{eq:oc-dynsys} can be adopted unchanged for our control methods. One major generalization, though, will be made in the context of multi-objective optimization (\secref{sec:moo}); there $\Gamma$ is replaced by a vector of independent cost functionals $\vec{\Gamma} = (\Gamma_1, \ldots, \Gamma_N)$. Another adjustment concerns the specialization of $\vec{\tilde{f}}$ and $\vec{u}$ for the particular control scheme considered here, as will be described next.

\begin{figure}[htpb]
    \centering
    \includegraphics[width=.4\textwidth]{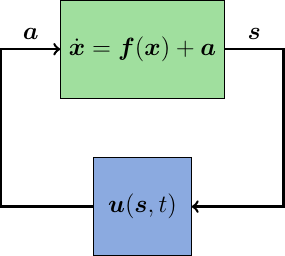}
    \caption{Sketch of the feedback control loop. The output of the dynamical system $\vec{x}$ is measured by sensors $\vec{s}$ which are used as input to the control function $\vec{u}$. The control function, in turn, acts on the system via actuators $\vec{a} = \vec{u}(\vec{s}, t)$ in order to achieve a desired state. (External disturbances which can be incorporated explicitly as additional inputs to the dynamical system and the control function are not shown here.) \label{fig:control-loop}}
\end{figure}

A feedback control scheme \cite{Gene2014FeedbackControl} is adopted in this work to implement the control. \figref{fig:control-loop} depicts the general architecture.

For this particular scheme Eq.~\eqref{eq:oc-dynsys} can be rewritten as:
\begin{align*}
    \dot{\vec{x}} = \vec{f}(\vec{x}, t) + \vec{a},\,\, \vec{x}(t_s) = \vec{x}_s,
\end{align*}
where the uncontrolled system $\dot{{\vec{x}}} = \vec{f}(\vec{x}, t)$ is controlled by an additive actuator term $\vec{a}$, and the control function now depends on sensor measurements, given by the output vector $\vec{s} \in \R^{n_s}$:
\begin{align*}
    \vec{a} = \vec{u}(\vec{s}, t).
\end{align*}
These measurements might be nonlinear functions of the state $\vec{x}$. For simplicity, external perturbations to the dynamic system are not considered here.

\subsection{Genetic Programming \label{sec:gp}}

\begin{figure}[htpb]
    \centering
    \includegraphics[width=.3\textwidth]{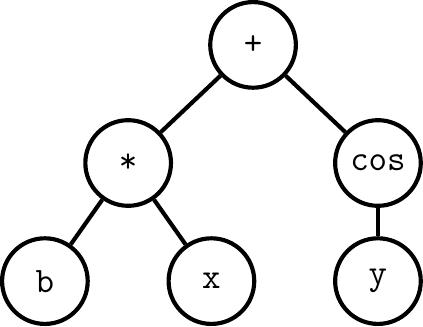}
    \caption{Tree representation of the mathematical expression $b \cdot x + \cos(y)$. The symbols \texttt{b}, \texttt{x}, and \texttt{y} are taken from the terminal set and make up the leaf nodes of the tree, whereas the symbols \texttt{*}, \texttt{+}, and \texttt{cos}, are symbols taken from the function set, they make up the internal nodes. \label{fig:gp_tree}}
\end{figure}

Genetic programming \cite{Koza1992GeneticProgramming,Cramer1985Representation} is an evolutionary algorithm in the class of meta-heuristic search techniques that promise global optimization. Similar to the genetic algorithm (GA), GP also uses the natural selection metaphor to evolve a set, or so-called population, of solutions or individuals using a cost-based selection mechanism. The evolution occurs over a number of iterations, called generations. GP differs from GA mainly in the representation of a solution. A solution in GP is generally represented using lists or expression trees. Expression trees are constructed from the elements of two predefined primitive sets: a function set consisting of mathematical operators and trigonometric functions, such as $\{$\texttt{+, -, *, cos, sin}$\}$, and a terminal set consisting of variables and constants, such as $\{$\texttt{ x, y, b}$\}$. Function symbols make up the internal nodes of a tree; and terminal symbols are used in the leaf nodes. For example, \figref{fig:gp_tree} shows the tree representation for the expression $b \cdot x + \cos(y)$. All elements of the tree are drawn from the aforementioned primitive sets: the variables and constants in the terminal set ($x$, $y$, and $b$) form the leaves of the tree and the mathematical symbols in the functional set ($\cdot$, $+$, and $\cos$) are used in forming the tree's internal nodes.

\begin{algorithm}[H]
\begin{algorithmic}
\Procedure{main}{}
\State $G_0 \leftarrow \text{random}(\lambda)$
\State $\text{evaluate}(G_0)$
\State $t \leftarrow 1$
\Repeat
\State $O_t \leftarrow \text{breed}(G_{t-1}, \lambda)$
\State $\text{evaluate}(O_t)$
\State $G_{t} \leftarrow \text{select}(O_t, G_{t-1}, \mu)$
\State $t \leftarrow t + 1$
\Until{$t > T \textbf{ or } G_t = \text{good}()$}
\EndProcedure
\end{algorithmic}
\caption{Top level description of a GP algorithm}
\label{alg:mumu}
\end{algorithm}

\begin{figure}[htpb]
    \centering
    \includegraphics[width=.8\textwidth]{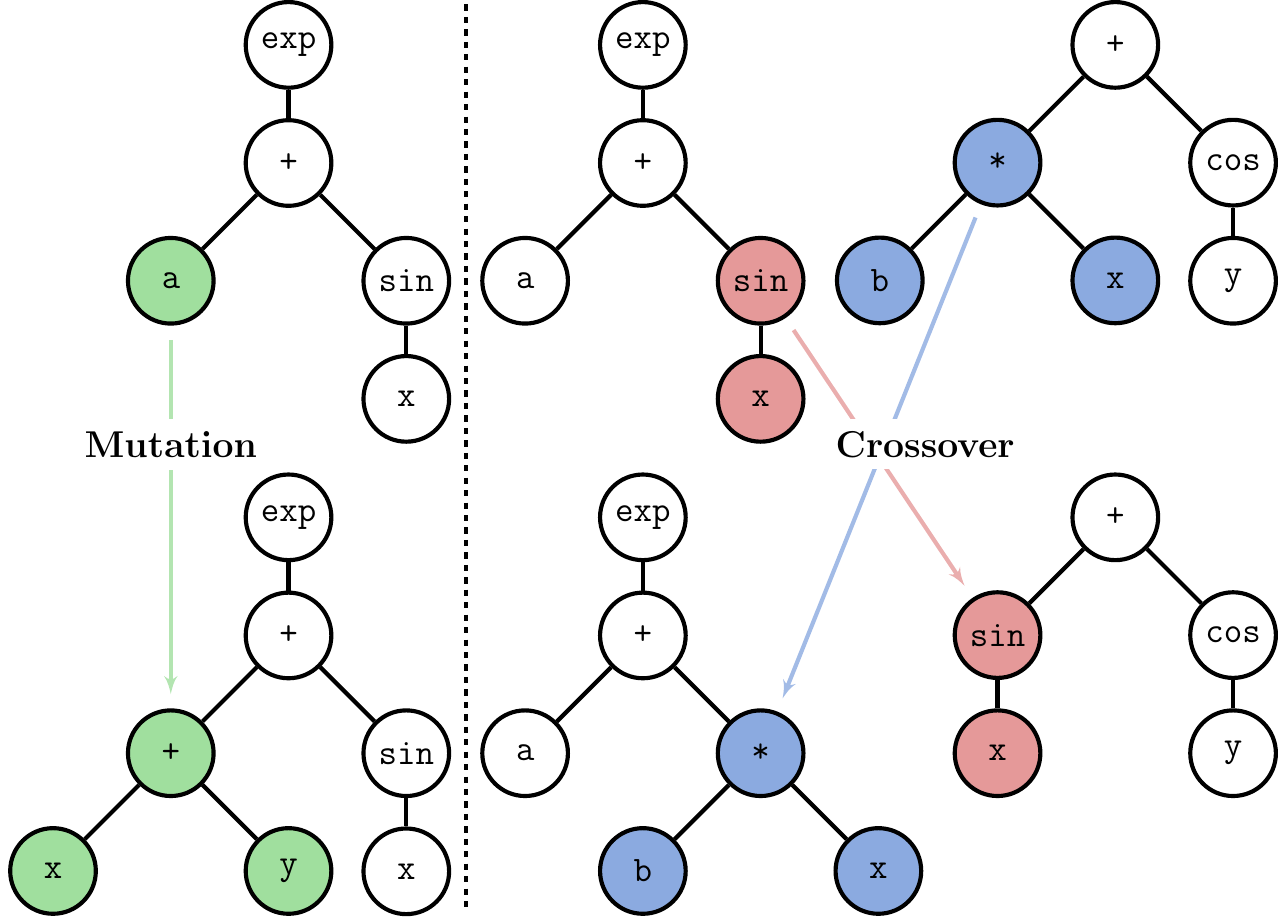}
    \caption{Breeding: Mutation and crossover operations on expression trees. \emph{Source:} Adapted from \cite{Quade2016Prediction}; used with permission. \label{fig:gp-mutation-and-crossover}}
\end{figure}

A standard GP algorithm, see Algorithm~\ref{alg:mumu}, begins by generating a population of random solutions $G_0$. A random variable length solution, with a given maximum tree depth, is generated by choosing operators, functions and variables from the two sets uniform randomly. Each solution is then evaluated in a given task, e.g., learning underlying relationship between a given set of variables. A cost is assigned to each solution based on its performance in solving the task, e.g. how closely a solution predicted the target function output. A new population of solutions $O_t$ is then generated by:
(i) probabilistically selecting parent solutions from the existing population using a cost-proportional selection mechanism, and
(ii) creating offspring or new solutions by applying recombination (or crossover) and variation (or mutation) operators (see \figref{fig:gp-mutation-and-crossover}).
This operation is repeated until a given number of solutions (a fixed population size) is reached. A closure property is always maintained to ensure only valid solutions are generated during both the initialization and breeding operations. An elitist approach is commonly used for improving the algorithm's convergence speed. This involves copying some of the high-performing parent solutions to the next-generation population $G_{t+1}$. The selection, evaluation and reproduction processes are repeated until a given stopping criteria is met, commonly a fixed number of maximum generations. For further details about GP operation, readers are referred to \cite{Poli2008FieldGuide,Yang2011Metaheuristics}.

The original motivation behind GP was to automatically produce computer programs, similar to a human strategy \cite{Koza1992GeneticProgramming}. However, since its inception, GP has been applied to various tasks, including the traditional machine learning \cite{Luke2013Metaheuristics} and optimization \cite{opt4ml} tasks. In the context of learning dynamical system models or control laws for dynamical systems, GP is a preferred choice for two main reasons: First, the expression tree representation used by GP is human interpretable and clearly provides an edge over the blackbox models learned by other computational approaches, such as artificial neural networks. Second, the GP solutions can be readily represented as mathematical equations and evaluated as control laws for dynamical systems.

\begin{figure}[htpb]
    \centering
    \includegraphics[width=.8\textwidth]{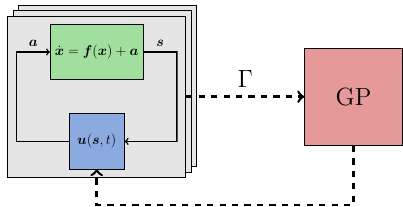}
    \caption{Sketch of the machine learning loop. Using an evolutionary metaphor, the GP algorithm generates a set of candidate control solutions $\vec{u}$, called the population. The candidate solutions are then evaluated in many realizations of the control loop; the performance in each iteration is rated via a cost functional $\Gamma$ and fed back as a cost index into the GP algorithm. The algorithm uses the performance rating to select the best solutions and evolve them into the next generation of candidate solutions. This learning loop repeats until at least one satisfactory control law is found (or other break conditions are met). \label{fig:learning-loop}}
\end{figure}

The application of GP to a general control problem (cf. \secref{sec:oc}), can be formulated as a learning and optimization task. That is, we would like to learn a control function to drive and keep a dynamical system in a desired state. This requires minimizing a cost function ($\vec{\Gamma}$), e.g., the difference between a given state in time and the desired state.
For most practical purposes, $\vec{\Gamma}$ can be expected to have complex properties including non-linearity, multi-modality, multi-variability and discontinuity. This poses a serious challenge to many traditional direct and gradient methods. In turn, meta-heuristic methods, such as GP, are suitable candidates  for this task. \figref{fig:learning-loop} depicts how GP-based dynamic controller is used within a feedback control loop, shown in \figref{fig:control-loop}.
\subsubsection{Multi-Objective Cost Evaluation\label{sec:moo}}
\begin{figure}[htpb]
    \centering
    \begin{overpic}[width=.9\textwidth]{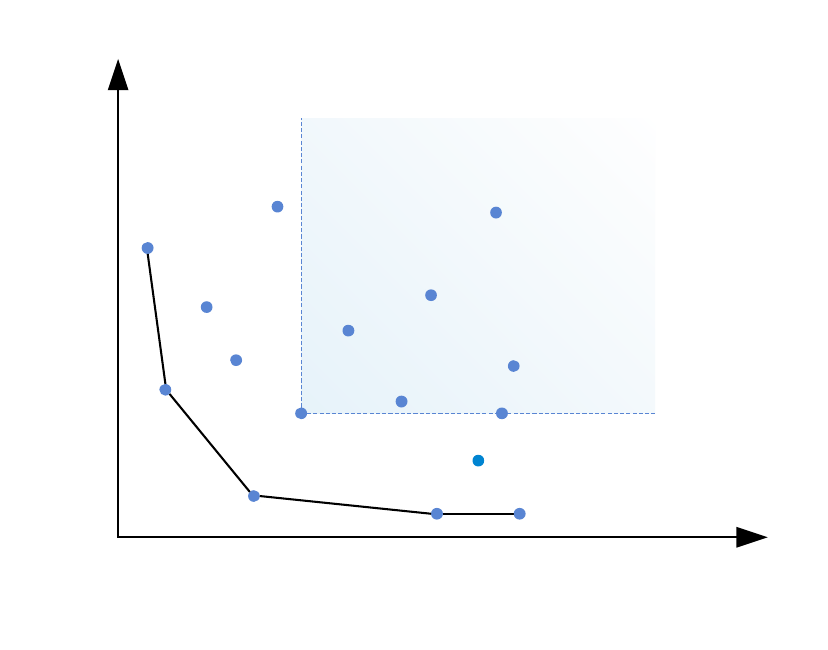}
        \put (34.5,25.5) {\textsf{A}}
        \put (84,9) {$\Gamma_1$}
        \put (9.5,62) {\rotatebox{90}{$\Gamma_2$}}
    \end{overpic}
    \caption{Pareto front (blue dots connected by a black line) of a set of candidate solutions evaluated with respect to two cost indexes $\Gamma_1$ and $\Gamma_2$. The Pareto-optimal solutions are non-dominated with respect to each other but dominate all other solutions. The region marked in light blue illustrates the notion of Pareto dominance: solutions contained within that region are dominated by a solution \textsf{A}. \label{fig:pareto-front}}
\end{figure}

Defining a good cost function is a key process in GP that matter, which determines the quality of a solution in the population.
A common method for cost assignment is to map solutions' performances to scalar values within a given range, e.g., $[0,1]$. This method simplifies the ranking procedure needed for selection and reproduction processes. However, it also limits the number of performance objectives that can be considered in the cost evaluation. A straight-forward way to address this concern is to apply a weighted sum method that in turn still allows mapping multiple performance objectives to scalar values. This type of methods not only require manual tuning of weights but also hide trade-off details between conflicting objectives. An alternative method to handle conflicting performance objectives in the design of cost functions is to use the concept of Pareto dominance \cite{fudenberg1991game}. According to this principle, a solution $x$ dominates another solution $y$, if $x$ performs better than $y$ in at least one of the multiple objectives or criteria considered and at least equal or better in all other objectives. This concept provides a convenient mechanism to consider multiple conflicting performance objectives simultaneously in ranking solutions based on their domination score. The solutions with the highest scores form the Pareto or efficient frontier. \figref{fig:pareto-front} provides an illustration of this concept. Several very successful evolutionary multi-objective optimization (EMO) algorithms are based on Pareto dominance \cite{Deb2000NSGAII,Knowles1999Pareto,Zitzler2001SPEA2}.

The standard GP is known to have a tendency to generate long and complicated solutions in order to exactly match, or overfit, performance target (optimal performance in our case). One way to address this issue is to design a cost function where both the performance (e.g., controller error in our case) and length of solutions are considered explicitly in determining the quality of a solution. Such a multi-objective cost mechanism allows introducing an explicit selection pressure in the evolutionary process and preferring smaller well-performing solutions over their longer counterparts \cite{Quade2016Prediction}. Following this line, a multi-objective cost evaluation method is adopted in our implementation of GP in this work. In specific, we adopt the mechanism used in NSGA-II \cite{Deb2000NSGAII} in our implementation of GP, which combines a non-dominated sorting mechanism with an Euclidean distance-based metric to promote solution diversity and spread or coverage of the entire Pareto front.
\subsubsection{Constant Optimization\label{sec:constopt}}
The learning of numeral constants, if desired, is treated similar to the learning of other terminal variables in the standard GP and an approximation may be learned by sampling over a given range. However, this approach can severely impair the convergence speed of GP as the search space essentially becomes infinite, especially for a continuous representation. Another approach, followed in this work, is to use a traditional parameter optimization algorithm, such as the Levenberg--Marquardt least squares method. By allowing designated symbolic constants $\vec{k} = (k_1, k_2, \ldots)$ in an expression tree, one can determine optimal values for these constants through parameter optimization. Thus, the calculation of the numeral constants becomes another optimization task, with
\begin{align*}
    \vec{k}^* = \underset{\vec{k}}{\text{argmin}} \, \Gamma(\vec{u}(\vec{s}, \vec{k})),
\end{align*}
and
\begin{align*}
    \Gamma = \Gamma(\vec{u}(\vec{s}, \vec{k}^*)),
\end{align*}
effectively introducing two combined layers of optimization.

To incorporate such a regression mechanism, the terminal set can be divided further into an argument set and a constant set, where the argument set contains all the terminal symbols that represent elements from the sensor vector $\vec{s}$, which are passed as arguments to the control function $\vec{u}$. The constant set, on the other hand, consists of all the designated constants representing elements from the constant vector $\vec{k}$. The actual construction of these sets depends on the dynamical system under consideration and the type of the control task. 

\subsubsection{GP setup\label{sec:gpsetup}}

This section gives a brief summary of the specific implementation details and the common parameters used in our experimental setup.
Hyperparameters have been chosen empirically such that they lead to plausible and interpretable results on the chosen set of examples. We did not optimize the hyperparameters for convergence.

All computer programs are developed in Python using open-source software packages. The implementation of the GP algorithm uses a customized version of the \textit{deap} module (Distributed Evolutionary Algorithms in Python) \cite{deap2012}. Particularly the routines for tree generation, selection, and breeding were adopted unchanged. Most of the numerical algorithms in use are provided by the \textit{numpy} and \textit{scipy} modules \cite{numpy2011,scipy2001}. Notably, constant optimization is conducted using the Levenberg-Marquardt least squares algorithm (\textit{scipy}) and numerical integration using the dopri5 solver (also \textit{scipy}). Random numbers are generated using the Mersenne Twister pseudo-random number generator provided by the \textit{random} module \cite{Matsumoto1998MersenneTwister}. Finally, the \textit{sympy} module is used for the simplification of symbolic mathematical expressions generated from the GP runs \cite{sympy2016}. The code can be found at \cite{markus_quade_2017_801819}.

\tabref{tab:setup} gives an overview of the methods and parameters used for the GP runs. Actual implementations can be found under the same name in the \textit{deap} module.

\begin{table}[h]
    \caption{General setup of the GP runs. \label{tab:setup}}
    \centering
    \begin{tabular}{ll}
        \toprule
        \addlinespace
        Function set & $\{+, -, \cdot, \sin, \cos, \exp\}$ \\  
        Population size & $500$ \\
        Max. generations & $20$\\
        MOO algorithm & NSGA-II \\
        \addlinespace
        Tree generation & \textit{halfandhalf} \\
        Min. height & 1 \\
        Max. height & 4 \\
        \addlinespace
        Selection & \textit{selTournament} \\
        Tournament size & $2$ \\
        \addlinespace
        Breeding & \textit{varOr} \\
        \addlinespace
        Recombination & \textit{cxOnePoint} \\
        Crossover probability & $0.5$ \\
        Crossover max. height & $20$ \\
        \addlinespace
        Mutation & \textit{mutUniform} \\
        Mutation probability & $0.2$ \\
        Mutation max. height & $20$ \\
        \addlinespace
        Constant optimization & \textit{leastsq} \\
        \addlinespace
        \bottomrule
    \end{tabular}
\end{table}

The function set is chosen such that the operators and functions are defined on $\R$. This allows for an easy evaluation and application of the expressions built from this set and prevents additional handling of singularities. This choice, though, restricts the number of possible solutions. In principle, the inclusion of other functions and operators, such as $\log$, $\sqrt{~}$, or $1/x$, is conceivable and would lead to a different space of potential solutions. Where possible, other GP parameter values are chosen according to the best practices used by the GP community; see \cite{Koza1992GeneticProgramming,Luke2013Metaheuristics,Poli2008FieldGuide}. For more involved cases a second stopping criterion is used where the learning is stopped when an error of the order of $10^{-5}$ is reached. Numerical integration is performed during cost assessment for all of the investigated dynamical systems. As mentioned before, dopri5 is used as solver \cite{press2007numerical}: This is an explicit Runge--Kutta method of order (4)5 with adaptive step size. If not otherwise stated, the maximum number of steps allowed during one call is set to $4000$, the relative tolerance to $10^{-6}$, and the absolute tolerance to $10^{-12}$. The pseudorandom number generator is seeded by a randomly selected unique seed for every GP run and never reinitialized during the same run. The specific seed used in an experiment will be explicitly stated in the corresponding setup description. Results obtained from one experiment can thus be duplicated when this same seed is reused in another run of the experiment.

\section{Control of independent dynamical systems\label{sec:examples}}

In order to establish a baseline, explain the working of the proposed method and determine its effectiveness, in this section we discuss the application of GP-based control methodology to a well-understood example, the harmonic oscillator.
It is forced to a fixed point and the resulting control laws are analyzed in detail. In addition, we investigated the application to the Lorenz system \cite{Lorenz1963}, and the results are found in the supplemental material.

\subsection{Harmonic Oscillator\label{sec:harmonic-oscillator}}

Consider a harmonic oscillator incorporated into the feedback control scheme discussed in \secref{sec:feedback-control}. The dynamical system reads:
\begin{align}
    \ddot{x} = -\omega_0^2 x + u(x, \dot{x}),\label{eqn:harmonic-oscillator}
\end{align}
with the particular system parameters defined in the left half of \tabref{tab:harmonic-oscillator-setup}. Ideal sensors are assumed to measure position and velocity, $\vec{s} = (x, \dot{x})$; thus, the explicit statement of the sensor function is omitted in the argument list of $u$.

The control target is to drive the harmonic oscillator toward a steady state resting position. This may be formulated into a cost function using the root mean squared error (RMSE) of the trajectory $x$ with reference to $0$, that is,
\begin{align*}
    \Gamma_1 := \text{RMSE}(x, 0) = \sqrt{\frac{1}{N} \sum\nolimits_{i=0}^{N-1}(x(t_i) - 0)^2}.
\end{align*}
Since this is a numerical experiment, the RMSE is formulated in a discrete form, with time steps $t_i = i \tfrac{T}{N}$ ($i=0,\ldots,N-1$), and an oscillation period $T = \tfrac{2\pi}{\omega_0}$. A time interval of twenty periods, as defined in \tabref{tab:harmonic-oscillator-setup}, is large enough to get a meaningful measurement of $\Gamma_1$. The number of discrete time steps, $n$, is chosen such that an accuracy of $50$ samples per period is achieved.

Further, as discussed in \secref{sec:moo}, the expression length is used as the second objective to bias GP learning towards smaller control laws:
\begin{align*}
    \Gamma_2 := \text{length}(u),
\end{align*}
which, corresponds to the number of nodes in the expression tree for $u$.

\begin{table}[htpb]
    \caption{Harmonic oscillator system setup.\label{tab:harmonic-oscillator-setup}}
    \centering
    \begin{tabular}{lllll}
        \toprule
        \multicolumn{2}{c}{Dynamic system} & \phantom{abc}& \multicolumn{2}{c}{GP} \\
        \cmidrule{1-2} \cmidrule{4-5}
        \addlinespace
        $\omega_0$     & $\exp(2)$                       && Cost functionals & $\text{RMSE}(x, 0)$ \\
        $x(t_0)$       & $\ln(4)$                        &&                  & $\text{Length}(u)$ \\
        $\dot{x}(t_0)$ & $0$                             && Argument set     & $\{x, \dot{x}\}$ \\
        $t_0,\,t_n$    & $0,\,20\tfrac{2\pi}{\omega_0}$  && Constant set     & $\{k\}$ \\
        $n$            & $1000$                          && Seed             & $1730327932332863820$ \\
        \addlinespace
        \bottomrule
    \end{tabular}
\end{table}

These two cost functions are listed as part of the GP setup in the right half of \tabref{tab:harmonic-oscillator-setup}. Additionally, the argument set and constant set are specified. In this case, the argument set consists of symbols representing position and velocity, the two quantities measured by $\vec{s}$; the constant set consists of a single constant, $k$, that is used to perform constant optimization. For easier readability, the notation does not distinguish between primitive symbols and variable names, that is, $x$ is used instead of \texttt{x}, and $\dot{x}$ instead of \texttt{x\_dot}. The seed from \tabref{tab:harmonic-oscillator-setup} is used to initialize the pseudorandom number generator of the GP algorithm and leads to the particular solutions presented next. Other relevant parameters are stated in the general GP setup, as described in \secref{sec:gpsetup}.

The Pareto solutions for this particular setup are presented in \tabref{tab:harmonic-oscillator-solutions} in ascending order sorted by $\Gamma_1$ (RMSE). Not surprisingly, more complex expressions tend to provide better control (in terms of lower RMSE). One can also observe multiple mathematically equivalent solutions in the Pareto set (e.g., the two expressions of length $6$). Although equivalent, these expressions are distinct as far as the internal representation is concerned and the GP algorithm treats them as independent solutions.

\begin{table}[htpb]
    \caption{Control of the harmonic oscillator: Pareto-front solutions.\label{tab:harmonic-oscillator-solutions}}
    \centering
    \begin{tabular}{ccll}
        \toprule
        RMSE & Length & \mc{Expression} & \mc{Constants} \\
        \midrule
        \addlinespace
        $0.043973$ & $10$ & $k \cdot (-k \cdot x + x - \dot{x})$ & $k = 151.907120$ \\
        $0.043976$ & $ 8$ & $k \cdot (-k \cdot x - \dot{x})    $ & $k = 151.232316$ \\
        $0.115862$ & $ 7$ & $\exp(-k \cdot x - \dot{x})        $ & $k = 3509.921747$ \\
        $0.123309$ & $ 6$ & $k \cdot (-x - \dot{x})            $ & $k = 8.545559$ \\
        $0.123309$ & $ 6$ & $-k \cdot (x + \dot{x})            $ & $k = 8.545559$ \\
        $0.123309$ & $ 5$ & $k \cdot (x + \dot{x})             $ & $k = -8.545254$ \\
        $0.127432$ & $ 3$ & $k \cdot \dot{x}                   $ & $k = -7.389051$ \\
        $0.241743$ & $ 2$ & $-\dot{x}                          $ &  \\
        $0.801177$ & $ 1$ & $k                                 $ & $k = 25.229759$ \\
        \addlinespace
        \bottomrule
        \addlinespace
        \addlinespace
    \end{tabular}
\end{table}

\begin{figure}[htpb]
    \centering
    \begin{overpic}[width=1.0\textwidth]{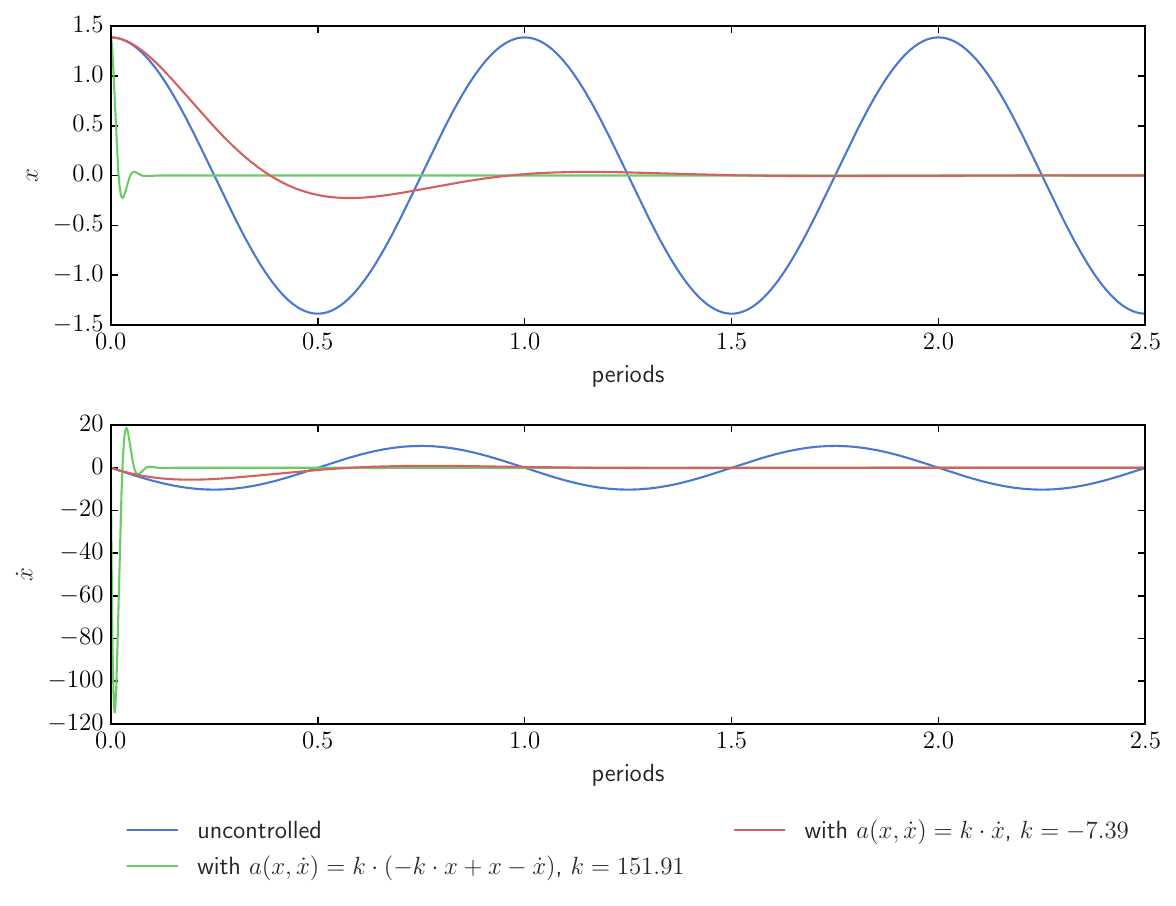}
        \put (0,74) {\textbf{(a)}}
        \put (0,40.1) {\textbf{(b)}}
    \end{overpic}
    \caption{Control of the harmonic oscillator. The trajectories of two candidate solutions, chosen from the Pareto front that drive the harmonic oscillator to zero are shown. The best solution regarding $\Gamma_1$ are shown in a green color. A simple yet moderately good solution with respect to $\Gamma_1$ is shown in red. For reference the uncontrolled system is shown in blue. \textbf{a} Position, \textbf{b} speed of the oscillators.\label{fig:harmonic_oscillator_phase_portrait}}
\end{figure}

\figref{fig:harmonic_oscillator_phase_portrait} shows the trajectories of two particular solutions from \tabref{tab:harmonic-oscillator-solutions} (first row and the third row from the bottom). Both look like underdamped cases of the damped harmonic oscillator system. We will analyze both solutions in more detail.

First consider $u(x, \dot{x}) = k (-k x + x - \dot{x})$. Inserting into the general Eq.~\eqref{eqn:harmonic-oscillator} for the controlled harmonic oscillator, one gets
\begin{align}
\begin{split}
    \ddot{x} &= -\omega_0^2 x + k (-k x + x - \dot{x}) \label{eqn:harmonic-oscillator-best}\\
             &= -(\omega_0^2 + k^2 - k) x - k \dot{x} \\
             &= -\tilde{\omega}_0^2 x - k \dot{x}
\end{split}
\end{align}
with $\tilde{\omega}_0^2 := \omega_0^2 + k^2 - k$. This is indeed the differential equation for the damped harmonic oscillator. Since $\omega_0^2 > 1$, it follows that $\tilde{\omega}_0^2 > 0$ and the condition for the underdamped case, $\tfrac{k^2}{4} < \tilde{\omega}_0^2$, is fulfilled for any $k \in \R$. Using the initial values from \tabref{tab:harmonic-oscillator-setup}, we get the particular solution
\begin{align}
    x(t) = e^{-\tfrac{k}{2} t + 2} \left( \cos(\omega t) + \frac{k \sin(\omega t)}{2 \omega} \right),\label{eqn:harmonic-oscillator-best-solution}
\end{align}
where $\omega^2 := \tilde{\omega}_0^2 - \tfrac{k^2}{4}$.

\begin{figure}[htpb]
    \centering
    \begin{overpic}[width=\textwidth]{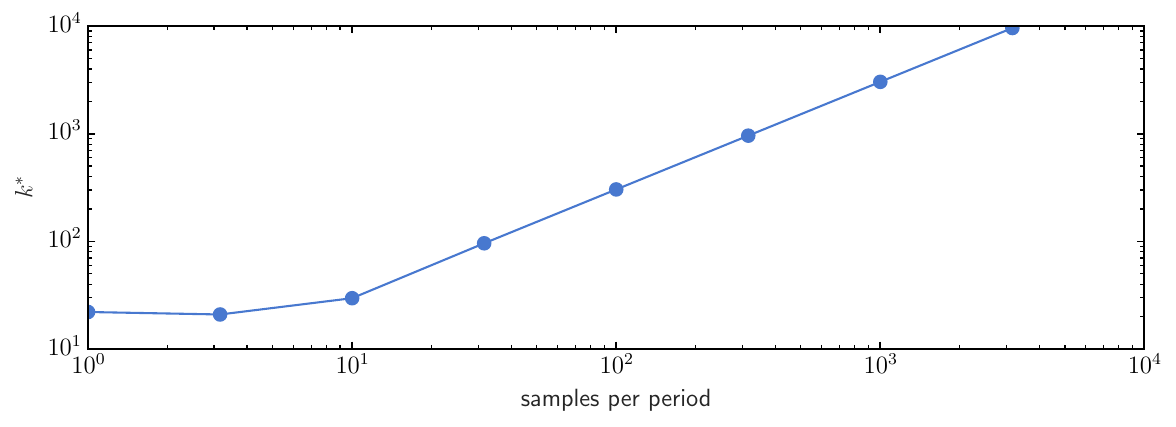}
    \end{overpic}
    \caption{Control of the harmonic oscillator. Optimal parameter $k^*$ for the control law $u(x, \dot{x}) = k \cdot (-k \cdot x + x - \dot{x})$ as determined by the least squares optimization using different sampling rates $f_s$. Starting at $f_s = 10$ the optimal value for $k$ increases linearly with the sampling rate, $k^* \sim f_s \sim 1 / \Delta t$.\label{fig:harmonic_oscillator_divergence_k}}
\end{figure}

The form of solution \eqref{eqn:harmonic-oscillator-best-solution} demands an answer to the specific value found for $k$: One would expect large values of $k \gg 151.9$ (up to the numeric floating point limit), since one of the goals of optimization is to drive the harmonic oscillator to zero, and the particular solution \eqref{eqn:harmonic-oscillator-best-solution} implies that $x(t) \xrightarrow{} 0$ as $k \rightarrow \infty$ ($t \in \R$). Why the least squares algorithm finds a considerably smaller value can be explained by the choice of discretization made here. \figref{fig:harmonic_oscillator_divergence_k} illustrates how the optimal value $k^*$ depends on the discretization of the finite time interval, more specifically the step size $\Delta t$. Since the result of numerical integration is restricted by the resolution of the time interval, starting at the initial position $x(0) = e^2$, the shortest time span possible for the trajectory to reach zero is $\Delta t$. Thus, setting an upper bound $k^*$ when optimizing for the RMSE of the whole trajectory with respect to zero: values larger $k^*$ would not improve the cost index any further.

\begin{figure}[htpb]
    \centering
    \begin{overpic}[width=\textwidth]{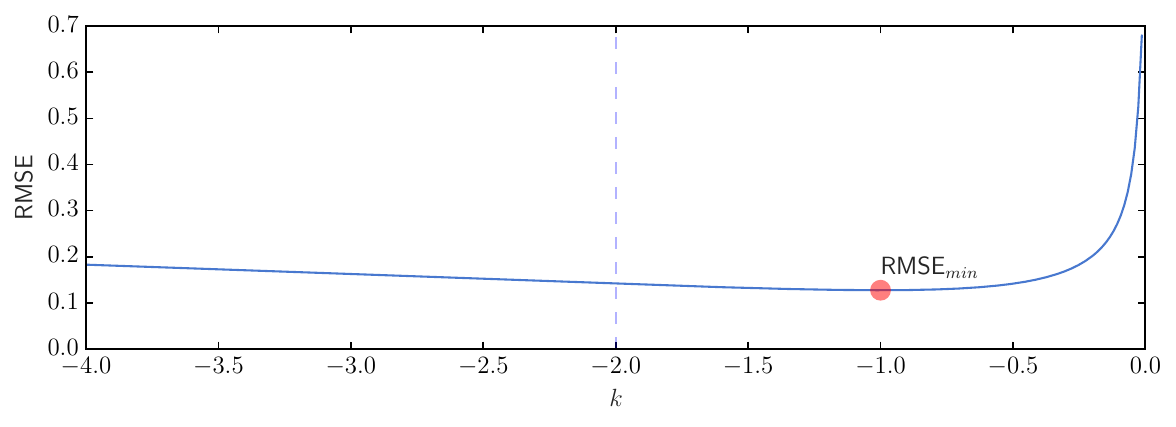}
    \end{overpic}
    \caption{Control of the harmonic oscillator. Minimum in RMSE with respect to $k$ for the control law $u(x, \dot{x}) = -k \cdot \dot{x}$. The minimum, at $k=-7.389 \approx -\omega_0$, lies in the underdamped regime. Solutions in the overdamped regime, left of the aperiodic borderline case (dashed line), result in strictly increasing RMSE as $k \rightarrow -\infty$. So do the diverging solutions for $k > 0$, as $k \rightarrow \infty$ (not shown).\label{fig:harmonic_oscillator_minimum_rmse}}
\end{figure}

As the second case, consider the solution $u(x, \dot{x}) = k \cdot \dot{x}$. Again inserting into \eqref{eqn:harmonic-oscillator}
\begin{align}
\begin{split}
    \ddot{x} &= -\omega_0^2 x + k \dot{x} \label{eqn:harmonic-oscillator-moderate},
\end{split}
\end{align}
one gets a damped harmonic oscillator system. The difference to the previous case \eqref{eqn:harmonic-oscillator-best-solution}, is that the coefficient of $x$ does not depend on $k$. This allows for a wider range of solutions that cover all regimes of the damped harmonic oscillator (i.e., overdamped, underdamped, and diverging case). A numerical analysis of the RMSE with respect to $k$ shows the presence of a single minimum in the underdamped regime; see \figref{fig:harmonic_oscillator_minimum_rmse}. The result $k^* = -\omega_0$ from constant optimization corresponds almost exactly to the minimum position, as would be expected.

\section{From Coupled Oscillators to Networks\label{sec:results}}

In this section, we extend the above study and demonstrate the application of GP-based control to networks of oscillators. As discussed in the introduction, such networks are used to model highly nonlinear complex systems, including the human brain. Given the latter as an application, the target structure is a hierarchical network. Nevertheless we want to systematically investigate the results of our method starting with a well-understood situation, namely two coupled oscillators. To step toward a network structure, we extend this first to a one-dimensional ring, or chain of coupled oscillators with periodic boundary conditions. Then we consider the two-dimensional analogon, the torus. Eventually, we study
a hierarchical network which has been proposed as a simplified model for the human brain \cite{Bullmore2009ComplexBrainNetworks}. The two coupled oscillators and the network results are discussed below; 1D and 2D periodic structures are discussed in supplemental material, since there is no essential new information in the results. Nevertheless one recognizes that results are consistent.

The aim, for all systems under consideration, is to control the synchronization behavior of the coupled oscillators. This can be done in two ways: starting from a synchronization regime and forcing the system into de-synchronization or vice versa, i.e., starting from a de-synchronized regime and forcing the system into synchronization. Both control goals are evaluated for the systems investigated.

\begin{figure}
    \centering
    \includegraphics[width=1.0\textwidth]{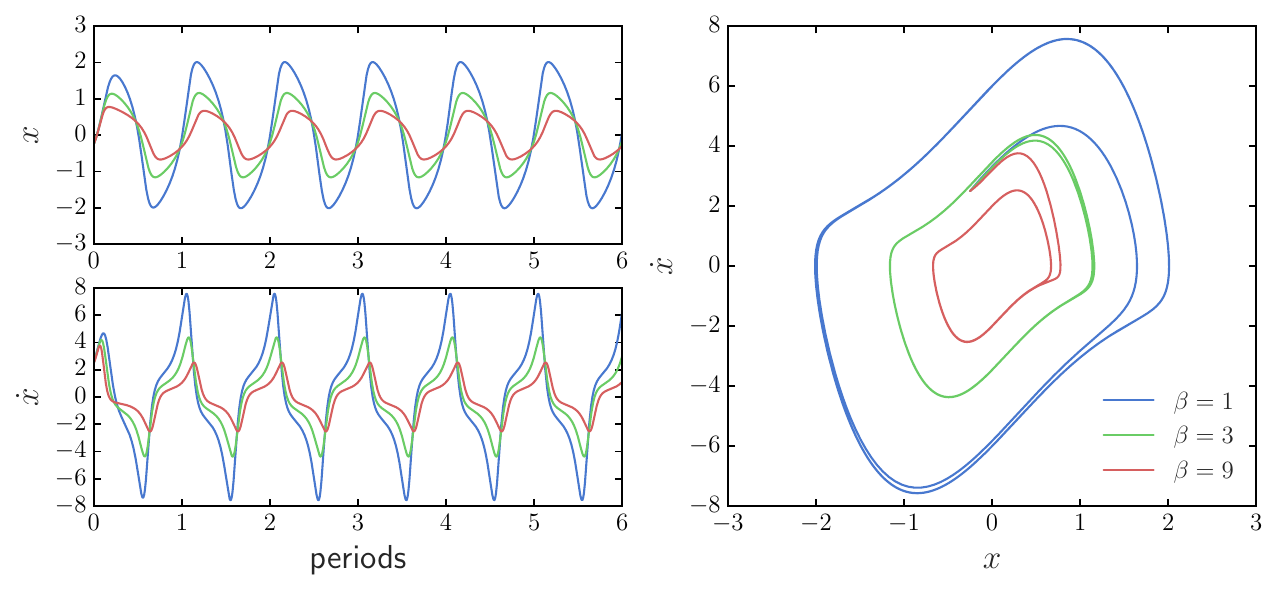}
    \caption{Single van der Pol oscillator (with parameters $\omega=e^2$, $\alpha=3$, and initial conditions $x(0) = -0.25$, $\dot{x}(0) = 2.5$.)\label{fig:van_der_pol}}
\end{figure}

The synchronization of dynamical systems is a well-known phenomenon exhibited by diverse ensembles of oscillators and oscillatory media \cite{Pikovsky2003Synchronization}. Here we will focus on a simple, but popular representative, the van der Pol oscillator, also used as a simple model for neurons. The van der Pol oscillator shows a nonlinear damping behavior governed by the following second-order differential equation:
\begin{align}
    \ddot{x} = - \omega^2 x + \alpha \dot{x} \left( 1 - \beta x^2 \right) =: f_{\text{vdP}}(x, \dot{x}),\label{eqn:van-der-pol}
\end{align}
where $x$ is the dynamical variable and $\omega$, $\alpha$, $\beta > 0$ are model parameters. The parameter $\omega$ is the characteristic frequency of the self-sustained oscillations, that is, the frequency at which the system tends to oscillate in the absence of any driving or damping force. The parameter $\alpha$ controls the non-linearity of the system. When $\alpha=0$, Eq.~\eqref{eqn:van-der-pol} becomes the harmonic oscillator. The damping parameter $\beta$ controls the dilation of the trajectory in the phase space. See \figref{fig:van_der_pol}.

A wide variety of coupling mechanisms exist. For hierarchical networks the investigation can be restricted to linearly coupled van der Pol oscillators in $x$ and $\dot{x}$, which can be described by a global coupling constant and two coupling matrices. An uncontrolled system of $N$ coupled van der Pol oscillators can be stated as follows:
\begin{align}
    \ddot{x}_i = f_{\text{vdP}}(x_i, \dot{x}_i) + c_1 \sum_{j=0}^{N-1} \kappa_{ij} x_j + c_2 \sum_{j=0}^{N-1} \varepsilon_{ij} \dot{x}_j \qquad (i = 0,\ldots,N-1)\label{eqn:coupled-van-der-pol}
\end{align}
with initial conditions
\begin{align*}
    x_{i}(t_0) = x_{i,0},\qquad \dot{x}_{i}(t_0) = \dot{x}_{i,0},
\end{align*}
where $c_{1,2}$ are the global coupling constants and $(\kappa_{ij})$ and $(\varepsilon_{ij})$ are the respective coupling matrices. This allows for several types of coupling such as direct, diffusive, and global coupling, or any other kind of network-like coupling. In the following experiments, we will use diffusive coupling in $\dot{x}_i$ for the topologies mentioned above. For GP, we use the same setup described above (\secref{sec:gpsetup}).

\subsection{Two Coupled Oscillators\label{sec:two_vdp}}

The simplest system showing synchronization is a system of two dissipatively coupled van der Pol oscillators \cite{Pikovsky2003Synchronization}:
\begin{align}
\begin{split}
   \ddot{x}_0 = - \omega_0^2 x_0 + \alpha \dot{x}_0 \left( 1 - \beta x_0^2 \right) + c \left( \dot{x}_1 - \dot{x}_0 \right),\\ \label{eqn:paired_vdp}
   \ddot{x}_1 = - \omega_1^2 x_1 + \alpha \dot{x}_1 \left( 1 - \beta x_1^2 \right) + c \left( \dot{x}_0 - \dot{x}_1 \right).
\end{split}
\end{align}
The coupling is restricted to $\dot{x}_i$, in which case the coupling constants from \eqref{eqn:coupled-van-der-pol} are set to $c_1 = 0$, $c_2 = c$, and the remaining coupling matrix reads as follows:
$ (\varepsilon_{ij}) =
\begin{bmatrix}
-1 &  1 \\
 1 & -1
\end{bmatrix}
$.

For the uncoupled oscillators, there are some parameter combinations $(\alpha, \beta)$ for which there exist stable limit cycles with characteristic frequencies $\omega_{0,1}$. If the two oscillators are coupled by a given coupling constant $c \neq 0$, as in \eqref{eqn:paired_vdp}, a range of frequencies with $\omega_0 \neq \omega_1$ emerge, where both oscillators effectively oscillate in a common mode. This range of frequencies is called the synchronization region. With the variation in the coupling constant, this region changes in width.

\begin{figure}
   \centering
   \includegraphics[width=1.0\textwidth]{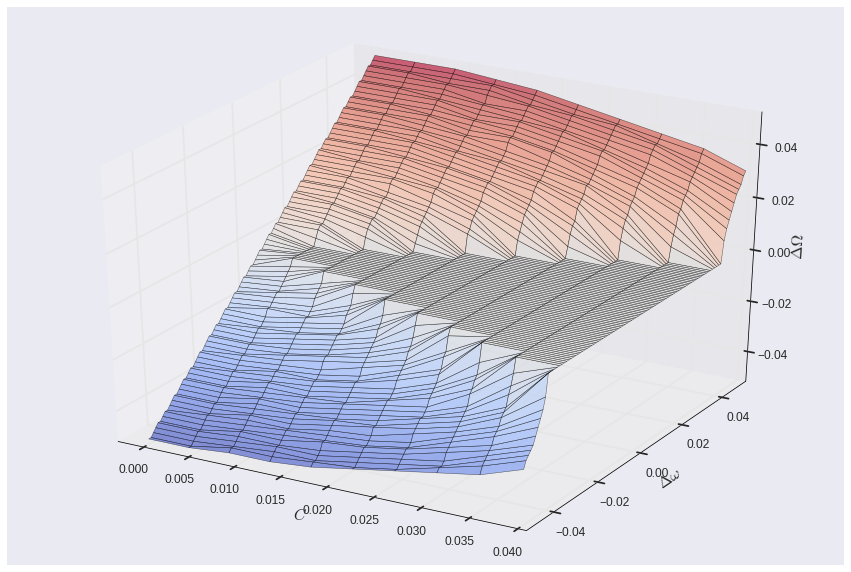}
   \caption{Synchronization plot of two coupled van der Pol oscillators with varying coupling strength $c$. The horizontal V-shaped plateau is referred to as the Arnold tongue; it represents regions of synchronization (The parameter set and initial conditions used are stated in the left part of \tabref{tab:two_vdp_sync_setup}).\label{fig:van_der_pol_sync_plot_full}}
\end{figure}

To illustrate this phenomenon, consider the particular parameter set $\alpha=0.1$, $\beta=1$, with fixed $\omega_0 = 1.386$ and varying $\omega_1$ in the range $[\omega_0  - 0.06,\, \omega_0 + 0.06]$. By plotting the observed frequency difference\footnote{The actual frequency $\Omega$ of an oscillator can be determined numerically by either taking the Fourier transform of the trajectory $x(t)$, or by counting the zero-crossings of $x(t) - \langle x(t) \rangle$.} $\Delta \Omega$, exhibited by the two oscillators, against the difference in their characteristic frequencies, $\Delta\omega := \omega_1 - \omega_0$, we can visualize the synchronization behavior of the system for a given coupling constant. See \figref{fig:van_der_pol_sync_plot_full}. Regions of synchronization show up as horizontal segments at $\Delta\Omega = 0$ (also, note the symmetry about $\Delta \omega = 0$). If this is done for several values of $c$ in the range $[0, 0.4]$, we can trace out the regions of synchronization. The result is a typical V-shaped plateau, the Arnold tongue.

\figref{fig:van_der_pol_sync_plot_full} shows the choice of appropriate parameters $\omega_1$ and $c$ to set up the system in different regimes for the purpose of control. The same approach is taken for all the experiments presented in this section.

Similar to the example systems in \secref{sec:examples} we add the control function $u$ to the equations \eqref{eqn:paired_vdp} of the uncontrolled system, yielding the following formulation:
\begin{align}
\begin{split}
   \ddot{x}_0 = - \omega_0^2 x_0 + \alpha \dot{x}_0 \left( 1 - \beta x_0^2 \right) + c \left( \dot{x}_1 - \dot{x}_0 \right) + u(\dot{\vec{x}}),\\ \label{eqn:paired_vdp_controlled}
   \ddot{x}_1 = - \omega_1^2 x_1 + \alpha \dot{x}_1 \left( 1 - \beta x_1^2 \right) + c \left( \dot{x}_0 - \dot{x}_1 \right) + u(\dot{\vec{x}}).
\end{split}
\end{align}
Here, $u$ is added as a global actuator term with equal influence on both oscillators; $u$ may depend on $\dot{x}_0$ and $\dot{x}_1$, summarized in vector notation as $\dot{\vec{x}} = (\dot{x}_0, \dot{x}_1)$.

\subsubsection{Forced Synchronization}

The system setup for forced synchronization of the two coupled van der Pol oscillators is presented in \tabref{tab:two_vdp_sync_setup}. The parameters $\omega_1$ and $c$ are chosen according to \figref{fig:van_der_pol_sync_plot_full}, such that the uncontrolled system follows a de-synchronization regime at a distance, $\Delta \omega$, approximately half the plateau from the closest synchronization point. The initial conditions are the same for both oscillators. The stopping criteria are chosen heuristically; in particular, the runtime was chosen from preliminary runs such that a typical control law can be found within that number of iterations.

The degree of de-synchronization is encompassed by the following cost functional:
\begin{align}
\begin{split}
    \Gamma_1 := |\Omega_0 - \Omega_1|.\label{eqn:two_vdp_sync_cost}
\end{split}
\end{align}
It measures the difference in observed frequencies exhibited by the two oscillators, with smaller differences reducing the cost on this objective.

As stated in the previous section, the actual frequencies $\Omega_0$ and $\Omega_1$ are numerically determined by counting zero crossings of the trajectory $x - \langle x \rangle$. This requires a careful choice of the time range $[t_0, t_n]$ of the observations, since the number of periods $N_P$ fitting into this interval determines an upper bound in absolute accuracy ($\sim\tfrac{1}{2N_P}$) of measuring $\Omega_0$, $\Omega_1$. Here, $N_P = 2000$ is chosen to yield an absolute accuracy well below $10^{-3}$ in the frequency range of interest.

\begin{table}[htpb]
    \caption{Two coupled oscillators: system setup for forced synchronization.\label{tab:two_vdp_sync_setup}}
    \centering
    \begin{tabular}{lllll}
        \toprule
        \multicolumn{2}{c}{Dynamic system} && \multicolumn{2}{c}{GP} \\
        \cmidrule{1-2} \cmidrule{4-5}
        \addlinespace
        $\omega_0$           & $\ln(4)$                         && Cost functionals & $|\Omega_0 - \Omega_1|$ \\
        $\omega_1$           & $\ln(4) + 0.04$                  &&                  & Length$(u)$ \\
        $\alpha,\,\beta,\,c$ & $0.1,\,1,\,0.022$                && Argument set     & $\{\dot{x}_0, \dot{x}_1\}$ \\
        $\vec{x}(t_0)$       & $(1,1)$                          && Constant set     & $\{k\}$ \\
        $\dot{\vec{x}}(t_0)$ & $(0,0)$                          && Seed             & $3464542173339676227$ \\
        $t_0,\,t_n$          & $0,\,2000\tfrac{2\pi}{\omega_0}$ && \\
        $n$                  & $40000$                          && \\
        \addlinespace
        \bottomrule
    \end{tabular}
\end{table}

Results from the GP run are presented in \tabref{tab:two_vdp_sync_results}. The algorithm stopped after one generation, providing six simple results optimally satisfying $\Gamma_1$. Since the equations \eqref{eqn:paired_vdp_controlled} are symmetric in $x_0$ and $x_1$, unsurprisingly so are the resulting control laws.

\begin{figure}[htpb]
    \centering
    \includegraphics[width=1.\textwidth]{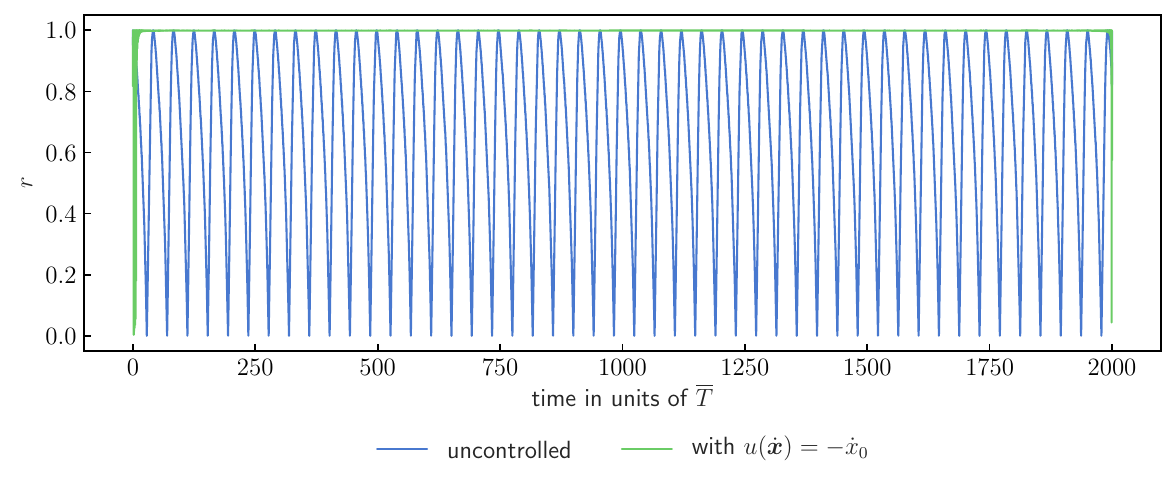}
    \caption{Two coupled oscillators: Kuramoto order parameter, $r$, for forced synchronization. Green: the controlled and blue: the uncontrolled system.\label{fig:two_vdp_sync_kuramoto}}
\end{figure}

To demonstrate the control effect \figref{fig:two_vdp_sync_kuramoto} shows the Kuramoto order parameter, $r$, representing phase coherence, plotted over time \cite{Kuramoto1975,Kuramoto1984}. The order parameter is defined by
\begin{align*}
    r &= \left| \frac{1}{N} \sum_{j=0}^{N-1} e^{i\varphi_j} \right|,
\end{align*}
with $\varphi_j$ the continuous phase  of the $j$th oscillator. It is computed from the analytical signal of the trajectory $x$ using the Hilbert transform, cf.\cite{Cohen1994TimeFrequencyAnalysis}. The controlled system completely synchronizes ($r \approx 1$) after a short initial period of de-synchronization, while the uncontrolled system exhibits an oscillating graph.

\begin{table}[htpb]
    \caption{Two coupled oscillators: Pareto-front solutions for forced synchronization.\label{tab:two_vdp_sync_results}}
    \centering
    \begin{tabular}{*{4}{c}}
        \toprule
        $|\Omega_0 - \Omega_1|$ & Length & \mc{Expression} & \\
        \midrule
        \addlinespace
        $0.0$ & $ 2$ & $\cos(\dot{x}_1)$ &  \\
        $0.0$ & $ 2$ & $\cos(\dot{x}_0)$ &  \\
        $0.0$ & $ 2$ & $-\dot{x}_0     $ &  \\
        $0.0$ & $ 2$ & $\sin(\dot{x}_1)$ &  \\
        $0.0$ & $ 2$ & $-\dot{x}_1     $ &  \\
        $0.0$ & $ 2$ & $\sin(\dot{x}_0)$ &  \\
        \addlinespace
        \bottomrule
    \end{tabular}
\end{table}

One recognizes that the algorithm favors  the least complex solution, using an asymmetric term, either damping in $x_1$ or $x_0$. We can analyze qualitatively these solutions,
 $u(\dot{\vec{x}})=-\dot{x}_0$ and $u(\dot{\vec{x}})=-\dot{x}_1$.
Let us take arbitrarily the $x_0$ term: Plugging in $u(\dot{\vec{x}}) = -\dot{x}_0$ into the first oscillator equation from \eqref{eqn:paired_vdp_controlled} we obtain
\begin{align*}
    \ddot{x}_0 &= - \omega_0^2 x_0 + \alpha \dot{x}_0 \left( 1 - \beta x_0^2 \right) + c \left( \dot{x}_1 - \dot{x}_0 \right) - \dot{x}_0 \\
               &= - \omega_0^2 x_0 + (\alpha - c - 1)\dot{x}_0 - \alpha \beta \dot{x}_0 x_0^2 + c \dot{x}_1 \\
               &= - \omega_0^2 x_0 + \tilde{\alpha}_1 \dot{x}_0 - \alpha \beta \dot{x}_0 x_0^2 + c \dot{x}_1 ,
\end{align*}
with $\tilde{\alpha}_1 = \alpha - c - 1$. Doing the same for the second oscillator equation
\begin{align*}
    \ddot{x}_1 &= - \omega_1^2 x_1 + \alpha \dot{x}_1 \left( 1 - \beta x_1^2 \right) + c \left( \dot{x}_0 - \dot{x}_1 \right) - \dot{x}_0 \\
               &= - \omega_1^2 x_1 + (c - 1)\dot{x}_0 - \alpha \beta \dot{x}_1 x_1^2 + (\alpha - c) \dot{x}_1 \\
               &= - \omega_1^2 x_1 + \tilde{\alpha}_2 \dot{x}_0 - \alpha \beta\dot{x}_1 x_1^2 + \tilde{c} \dot{x}_1 ,
\end{align*}
with $\tilde{\alpha}_2 = c - 1$ and $\tilde{c} = \alpha - c$, one gets a similar solution in terms of the dominating driving component $\dot{x}_0$. Since $\tilde{\alpha}_1 \approx \tilde{\alpha}_2$ and $\tilde{\alpha}_1 \gg \tilde{c},c$, the oscillators behave almost identically and, thus, are synchronized. Analogous observations apply when analyzing the second control law, $u(\dot{\vec{x}}) = -\dot{x}_1$.

\subsubsection{Forced De-Synchronization}

The system setup for forced de-synchronization is given in \tabref{tab:two_vdp_desync_setup}. The parameters $\omega_1$ and $c$ are, again, chosen according to \figref{fig:van_der_pol_sync_plot_full}, this time, such that the uncontrolled system follows a synchronization regime well inside the plateau. The measure for the degree of synchronization is reciprocal to the previous case
\begin{align}
\begin{split}
    \Gamma_1 := \exp(-|\Omega_0 - \Omega_1|).\label{eqn:two_vdp_desync_cost}
\end{split}
\end{align}
All the remaining parameters are the same as for forced synchronization (\tabref{tab:two_vdp_sync_setup}).

\begin{table}[htpb]
    \caption{Two coupled oscillators: System setup for forced de-synchronization.\label{tab:two_vdp_desync_setup}}
    \centering
    \begin{tabular}{lllll}
        \toprule
        \multicolumn{2}{c}{Dynamic system} && \multicolumn{2}{c}{GP} \\
        \cmidrule{1-2} \cmidrule{4-5}
        \addlinespace
        $\omega_0$           & $\ln(4)$                         && Cost functionals & $\exp(-|\Omega_0 - \Omega_1|)$ \\
        $\omega_1$           & $\ln(4) + 0.015$                 &&                  & Length$(u)$ \\
        $\alpha,\,\beta,\,c$ & $0.1,\,1,\,0.022$                && Argument set     & $\{\dot{x}_0, \dot{x}_1\}$ \\
        $\vec{x}(t_0)$       & $(1,1)$                          && Constant set     & $\{k\}$ \\
        $\dot{\vec{x}}(t_0)$ & $(0,0)$                          && Seed             & $2590675513212712687$ \\
        $t_0,\,t_n$          & $0,\,2000\tfrac{2\pi}{\omega_0}$ && \\
        $n$                  & $40000$                          && \\
        \addlinespace
        \bottomrule
    \end{tabular}
\end{table}

\begin{figure}[htpb]
    \centering
    \includegraphics[width=1.\textwidth]{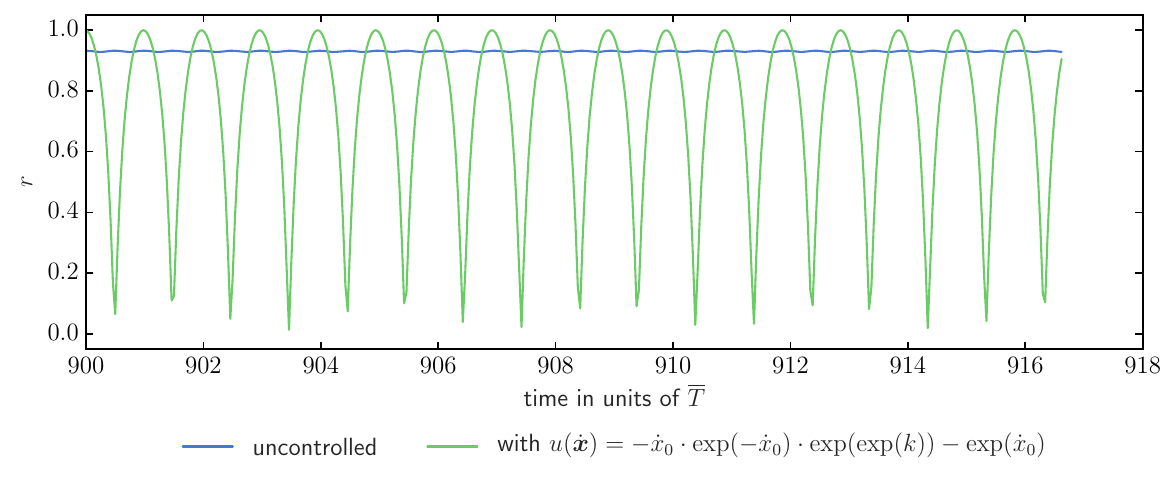}
    \caption{Two coupled oscillators: Kuramoto order parameter, $r$, for forced de-synchronization. Green: the controlled and blue: the uncontrolled system. The horizontal axis is scaled to a limited time window in order to make the oscillations visible.\label{fig:two_vdp_desync_kuramoto}}
\end{figure}

Results from the GP run are shown in \tabref{tab:two_vdp_desync_results}. Two aspects of the results indicate that de-synchronizing the pair of oscillators is a more demanding task: First, the GP algorithm comes up with increasingly long expressions to achieve improvements in $\Gamma_1$; second, constant optimization seems to fail in all cases where a constant is present. (This is expressed by a value $k=1$, which corresponds to the initial guess of the optimization procedure.) Still, the oscillating Kuramoto parameter, $r$, of the controlled system in \figref{fig:two_vdp_desync_kuramoto} shows that the best solution with respect to $\Gamma_1$ performs well in de-synchronizing the oscillators.

The control law $u(\dot{\vec{x}}) = -\dot{x}_0 \cdot \exp(\exp(k) + \cos(k)) = -\tilde{k} \dot{x}_0$, with $\tilde{k}\approx 26$, is almost the same as the solution analyzed in the previous subsection, but with a different coefficient. This is at first sight counterintuitive, but can be explained roughly by the non-uniqueness we provoke with our cost function: To force synchronization, we require only that the phase difference is small (close to zero). This is achieved by the added damping term. The very strong damping brings the two oscillators basically to zero so fast that the mutual coupling does not play a role and the phase difference is free.
To bring the oscillators from de-synchronization to synchronization is achieved by a different mechanism; the damping is moderate such that excess energy is dissipated and the oscillators are in the right regime to synchronize. The detailed analysis of the dynamics is subject of ongoing work, where we analyze the bifurcations occurring using AUTO. Preliminary results affirm that the interpretation given here is correct.

One would expect the GP algorithm to directly generate the simpler --- thus better suitable--- solution $u(\dot{\vec{x}}) = - k \dot{x}_0$, with $k = 26$. One possible explanation why this is not immediately found might lie in the constant optimization algorithm: On failure, the least squares algorithm returns the result of the last internal iteration. This return value might be an entirely inadequate value for $k$, which, in turn, could lead to an exploding cost index $\Gamma_1$ when integrating the dynamic system \eqref{eqn:paired_vdp_controlled}, hence disqualifying the corresponding solution.

\begin{table}[htpb]
\caption{Two coupled oscillators: Pareto-front solutions for forced de-synchronization.\label{tab:two_vdp_desync_results}}
\centering
\begin{tabular}{ccll}
    \toprule
    $\exp(-|\Omega_0 - \Omega_1|)$ & Length & \mc{Expression} & \mc{Constant} \\
    \midrule
    \addlinespace
    $0.248$ & $ 9$ & $-\dot{x}_0 \cdot \exp(\exp(k) + \cos(k))     $ & $k = 1$ \\
    $0.258$ & $ 7$ & $\cos(\exp(\dot{x}_1 + \cos(\cos(\dot{x}_0))))$ &  \\
    $0.875$ & $ 4$ & $\cos(\exp(\exp(\dot{x}_0)))                  $ &  \\
    $0.912$ & $ 3$ & $\sin(\exp(\dot{x}_0))                        $ &  \\
    $0.999$ & $ 2$ & $\exp(k)                                      $ & $k = 1$ \\
    $1.000$ & $ 1$ & $\dot{x}_1                                    $ &  \\
    $1.000$ & $ 1$ & $\dot{x}_0                                    $ &  \\
    $1.000$ & $ 1$ & $k                                            $ & $k = 1$ \\
    \\
    $0.247672$ & $13$ & $-\dot{x}_0 \cdot \exp(-\dot{x}_0) \cdot \exp(\exp(k)) - \exp(\dot{x}_0)$ & $k = 1.000000$ \\
    $0.386623$ & $12$ & $-\exp(\dot{x}_0) - \cos(\exp(\exp(k)) \cdot \exp(\sin(\dot{x}_0)))     $ & $k = 1.000000$ \\
    $0.398873$ & $11$ & $\sin(\exp(\exp(\dot{x}_0) + \cos(k)) \cdot \exp(\sin(\dot{x}_0)))      $ & $k = 1.000000$ \\
    $0.606677$ & $ 8$ & $\sin(\exp(\exp(k)) \cdot \exp(\sin(\dot{x}_0)))                        $ & $k = 1.000000$ \\
    $0.648420$ & $ 7$ & $\sin(\exp(\exp(\dot{x}_0) + \cos(k)))                                  $ & $k = 1.000000$ \\
    $0.806083$ & $ 6$ & $-\exp(\dot{x}_0) - \cos(k)                                             $ & $k = 1.000000$ \\
    $0.911933$ & $ 3$ & $\sin(\exp(\dot{x}_0))                                                  $ & $k = 1.000000$ \\
    $0.998615$ & $ 2$ & $\exp(k)                                                                $ & $k = 1.000000$ \\
    $1.000000$ & $ 1$ & $\dot{x}_1                                                              $ & $k = 1.000000$ \\
    $1.000000$ & $ 1$ & $\dot{x}_0                                                              $ & $k = 1.000000$ \\
    $1.000000$ & $ 1$ & $k                                                                      $ & $k = 1.000000$ \\
    \addlinespace
    \bottomrule
\end{tabular}
\end{table}

\subsection{Hierarchical Network}

\begin{figure}[h]
    \centering
    \includegraphics[width=\textwidth]{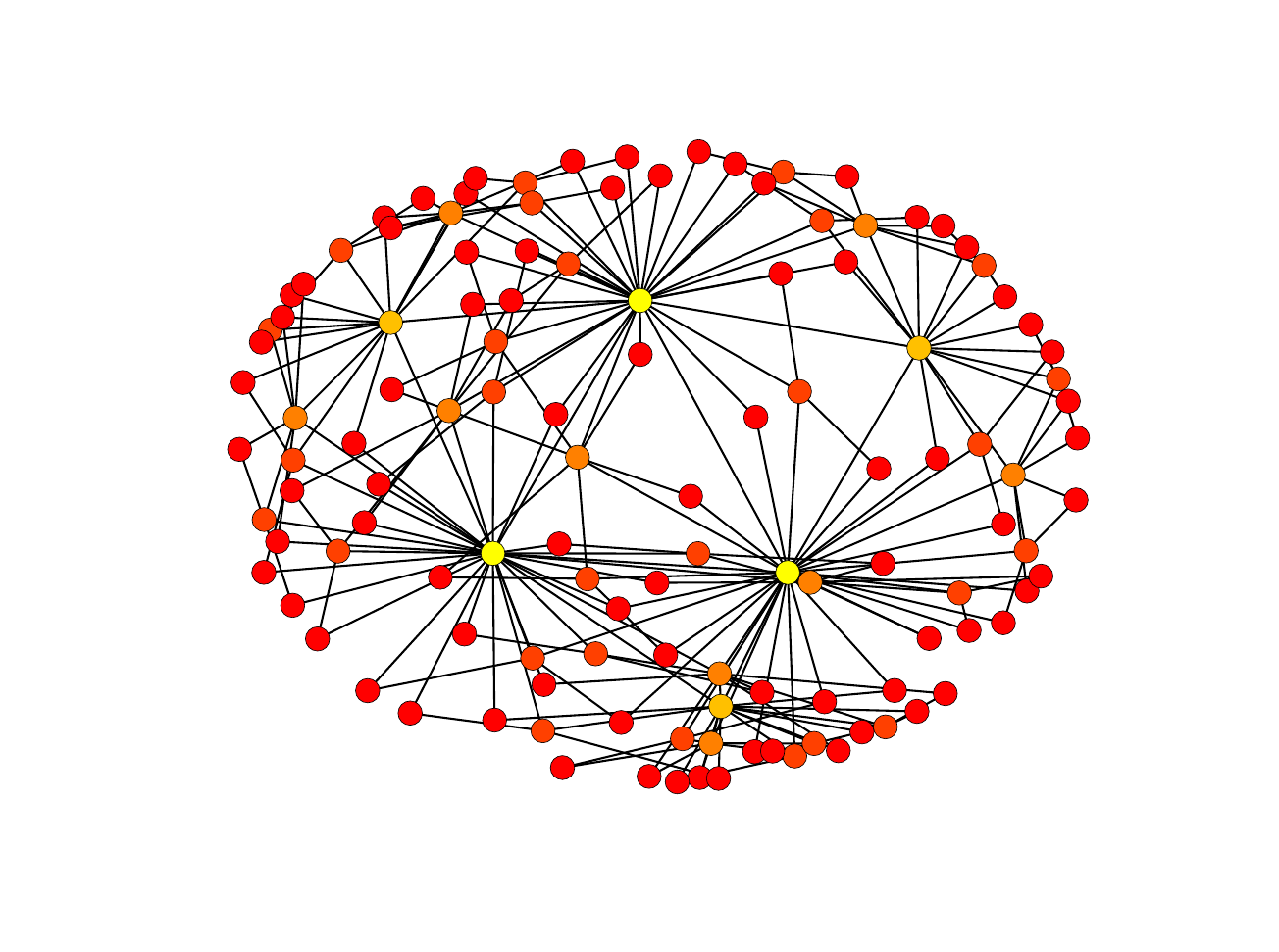}
    \caption{Hierarchical network: Dorogovtsev--Goltsev--Mendes topology of generation five. Starting out from two connected nodes at generation 0, one new node is added in between every existing pair of nodes per generation. Hence, the node degree at generation $n>0$ ranges from $2^1,\ldots,2^n$. The degree distribution, that is, the fraction $P(k)$ of nodes in the network having $k$ connections to other nodes goes for large values of $k$ as $P(k) \sim k^{-2}$ (Nodes are color-coded by node degree: yellow: 32, light orange: 16, orange: 8, dark orange: 4, red: 2).\label{fig:dorogovtsev_topology}}
\end{figure}

In a last step the dynamic system is extended to a set of van der Pol oscillators connected in a scale-free network topology. A scale-free network is a hierarchical network whose degree distribution follows a power law, at least asymptotically. There exists a large variety of possible models for creating networks which are able to reproduce the unique properties of the scale-free topology. One simple model, resorted to here, is the Dorogovtsev--Goltsev--Mendes model. It is used to produce a network of $N = 123$ nodes as depicted in \figref{fig:dorogovtsev_topology}.

The van der Pol oscillators are indexed in descending order by their corresponding node degree. For example, the three yellow nodes of degree 32 (highest) in \figref{fig:dorogovtsev_topology} are labeled $i=0,1,2$, the light orange nodes of the next lowest degree 16, $i=3,4,5$, and so on. The particular order of nodes of the same degree is not important due to the symmetry of the network.

``Sensors'' are placed on the oscillators labeled $i = 0,\ldots,11$ measuring $\dot{x}_i$. This incorporates all nodes with a node degree of 32 or 16, and five nodes with a node degree of 8. Hence, the control function $u$ can potentially make use of measurements from nodes at central points of the topology. Control, on the other hand, is exerted indiscriminately on all nodes of the system

\subsubsection{Forced Synchronization}

\tabref{tab:dorogovtsev_vdp_sync_setup} shows the setup. As in the related cases before, the particular parameter set chosen puts the uncontrolled system in a de-synchronization regime. Sensors are placed on the oscillators labeled $i = 0,\ldots,11$ measuring $\dot{x}_i$. This incorporates all nodes with a node degree of 32 or 16, and five nodes with a node degree of 8. Hence, the control function $u$ can potentially make use of measurements from nodes at central points of the topology. Control, on the other hand, is excerted indiscriminately on all nodes of the system
\begin{align*}
    \ddot{x}_i = f_{\text{vdP}}(x_i) + c \sum_{j=1}^{N} \varepsilon_{ij} \dot{x}_j + u(x_0,\ldots,x_{11}) \qquad (i = 0,\ldots,N-1),
\end{align*}
with $N=123$. Constant optimization is performed on a single constant $k$.

\begin{table}[htpb]
    \caption{Oscillators in a hierarchical network: system setup for forced synchronization.\label{tab:dorogovtsev_vdp_sync_setup}}
    \centering
    \begin{tabular}{lllll}
        \toprule
        \multicolumn{2}{c}{Dynamic system} && \multicolumn{2}{c}{GP} \\
        \cmidrule{1-2} \cmidrule{4-5}
        \addlinespace
        $\alpha,\,\beta,\,c$                 & $0.1,\,1,\,5.6 \cdot 10^{-2}$                && Cost functionals & std$(\vec{\Omega})$ \\
        $\overline{\omega}$, $\Delta \omega$ & $\ln(4)$, $8 \cdot 10^{-2}$                  &&                  & Length$(u)$ \\
        $ \omega_i$                          & linspace$(\overline{\omega} - \Delta \omega,
                                                         \overline{\omega} + \Delta \omega,
                                                         123)$                              && Argument set     & $\{\dot{x}_i\}_{i=0,\ldots,11}$ \\
        $\vec{x}_i(t_0)$                     & $1$ $(i=0,\ldots,122)$                       && Constant set     & $\{k\}$ \\
        $\dot{\vec{x}}_i(t_0)$               & $0$ $(i=0,\ldots,122)$                       && Seed             & $5925327490976859669$ \\
        $t_0,\,t_n$                          & $0,\,2000\tfrac{2\pi}{\overline{\omega}}$    && \\
        $n$                                  & $40000$                                      && \\
        \addlinespace
        \bottomrule
    \end{tabular}
\end{table}

\begin{figure}[htpb]
    \begin{overpic}[width=1.0\textwidth]{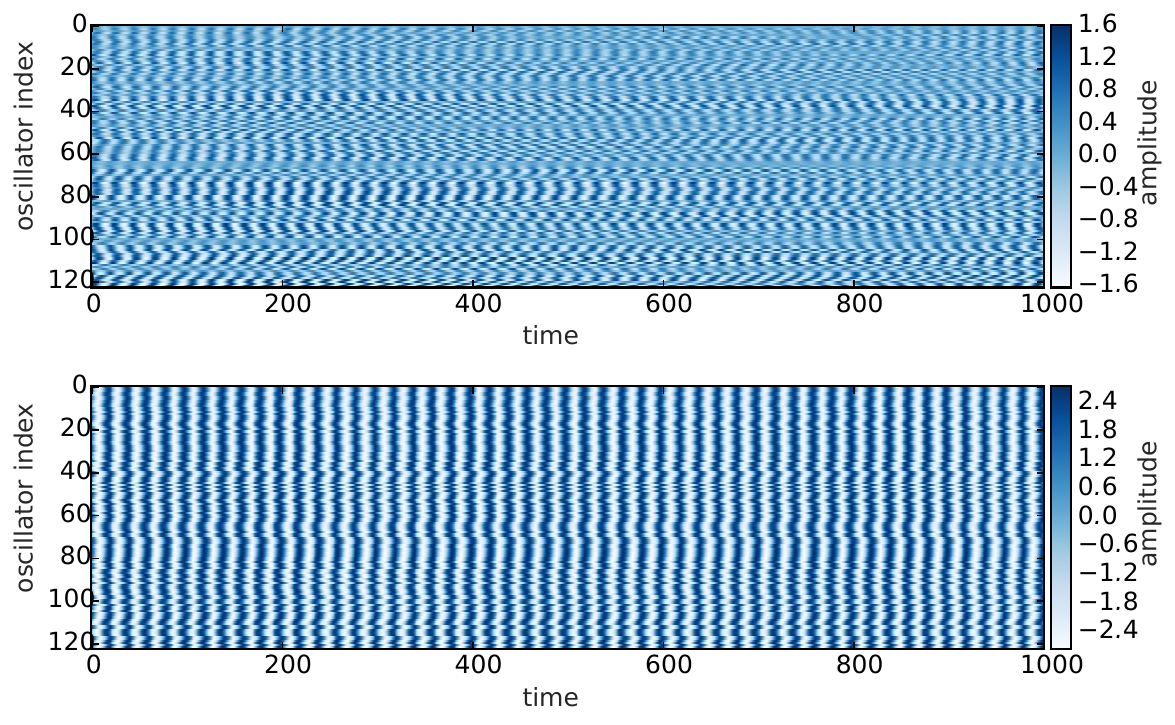}
        \put (0,59.5) {\textbf{(a)}}
        \put (0,29.0) {\textbf{(b)}}
    \end{overpic}
    \caption{Oscillators in a hierarchical network. \textbf{a} uncontrolled system; \textbf{b} Pareto-front solution, $u(\dot{\vec{x}})=\sin(\dot{x}_7)$, for forced synchronization.\label{fig:dorogovtsev_vdp_sync_heatmap}}
\end{figure}

\begin{figure}[htpb]
    \includegraphics[width=1.\textwidth]{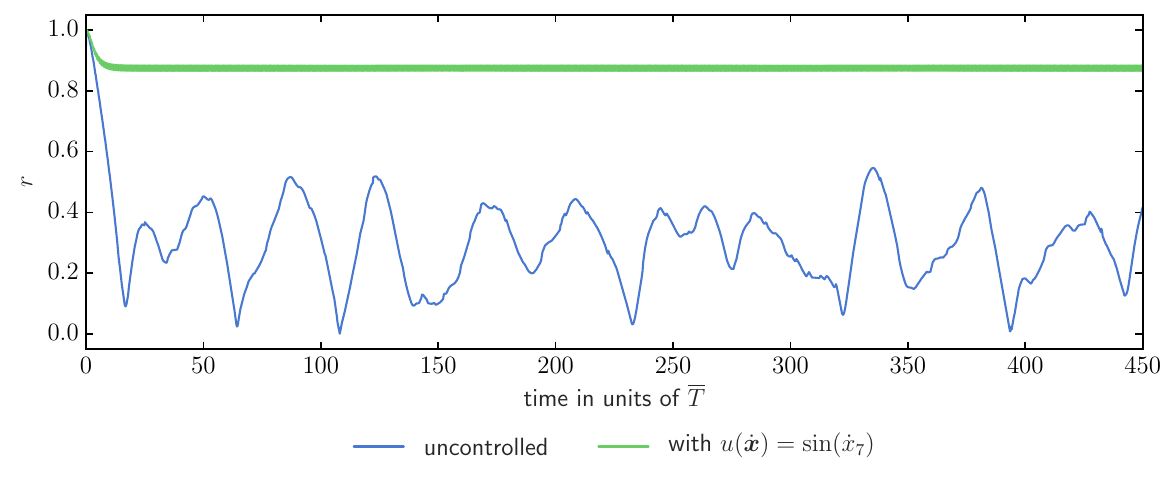}
    \caption{Oscillators in a hierarchical network. Kuramoto order parameter $r$ for the uncontrolled system (blue) and the system controlled by the best solution, with respect to $\Gamma_1$ (green).\label{fig:dorogovtsev_vdp_sync_kuramoto}}
\end{figure}

The GP run stopped after one generation with a Pareto front consisting of a single optimal result. See \tabref{tab:dorogovtsev_vdp_sync_results}. The control law found, is again, sinusoidal in nature. It uses the input from node $i=7$, which is of node degree eight, i.e., on an intermediate level in the topology. \figref{fig:dorogovtsev_vdp_sync_heatmap} shows that the highly distorted phases from the uncontrolled system can be partially aligned. Frequencies are approximately matched and amplitudes are amplified by a factor $\times 1.5$ with respect to the uncontrolled system. A plot of the Kuramoto order parameter $r$ in \figref{fig:dorogovtsev_vdp_sync_kuramoto} indicates that there is a small variation among the phases in the controlled system; hence, a perfect phase lock is not achieved.

\begin{table}[htpb]
    \caption{Oscillators in a hierarchical network: Pareto-front solutions for forced synchronization.\label{tab:dorogovtsev_vdp_sync_results}}
    \centering
    \begin{tabular}{ccll}
        \toprule
        std$(\vec{\Omega})$ & length & \mc{expression} \\ 
        \midrule
        \addlinespace
        $0.0$ & $ 2$ & $\sin(\dot{x}_7)$ \\
        \addlinespace
        \bottomrule
    \end{tabular}
\end{table}

\subsubsection{Forced De-Synchronization}

For the case of forced de-synchronization, the system parameters are set as in \tabref{tab:dorogovtsev_vdp_desync_setup}. The network structure stays as in the previous section.

\begin{table}[htpb]
    \caption{Oscillators in a hierarchical network: system setup for forced de-synchronization.\label{tab:dorogovtsev_vdp_desync_setup}}
    \centering
    \begin{tabular}{lllll}
        \toprule
        \multicolumn{2}{c}{Dynamic system} && \multicolumn{2}{c}{GP} \\
        \cmidrule{1-2} \cmidrule{4-5}
        \addlinespace
        $\alpha,\,\beta,\,c$                 & $0.1,\,1,\,5.6 \cdot 10^{-2}$                && Cost functionals & $\exp($std$(\vec{\Omega}))$ \\
        $\overline{\omega}$, $\Delta \omega$ & $\ln(4)$, $2 \cdot 10^{-2}$                  &&                  & Length$(u)$ \\
        $ \omega_i$                          & linspace$(\overline{\omega} - \Delta \omega,
                                                         \overline{\omega} + \Delta \omega,
                                                         123)$                              && Argument set     & $\{\dot{x}_i\}_{i=0,\ldots,11}$ \\
        $\vec{x}_i(t_0)$                     & $1$ $(i=0,\ldots,122)$                       && Constant set     & $\{k\}$ \\
        $\dot{\vec{x}}_i(t_0)$               & $0$ $(i=0,\ldots,122)$                       && Seed             & $8797055239111497159$ \\
        $t_0,\,t_n$                          & $0,\,2000\tfrac{2\pi}{\overline{\omega}}$    && \\
        $n$                                  & $40000$                                      && \\
        \addlinespace
        \bottomrule
    \end{tabular}
\end{table}

The previous sections showed increasingly complex control from two oscillators, a ring of oscillators, to a torus. One might expect an even more complex control for a hierarchical network. Indeed, in \tabref{tab:dorogovtsev_vdp_desync_result} the best control laws are complicated combinations of sine, multiplication and exponential terms. As explained above, the lowest indices indicate highest node degree. We observe that our algorithm puts control on nodes with high degree - the hubs. This is perfectly logical: If the hubs are de-synchronized, the whole system is desynchronized. It may still be that the hubs and their connected subnet are synchronized. This is not contained in our objective, and again, we find that our machine learns exactly the way it is told (Figs \ref{fig:dorogovtsev_vdp_sync_kuramoto}, \ref{fig:dorogovtsev_vdp_sync_heatmap}). If we want to control global de-synchronization in each single oscillator, we need to design the cost function with more care!

\begin{table}[htpb]
    \caption{Oscillators in a hierarchical network: best solutions for forced de-synchronization.\label{tab:dorogovtsev_vdp_desync_result}}
    \centering
    \begin{tabularx}{\textwidth}{ccXl}
        \toprule
        Synchronicity & Length & \mc{Expression} & \mc{Constant} \\
        \midrule
        \addlinespace
        $0.722$ & $17$ & $-(-\dot{x}_3 + (- \dot{x}_8)) + (\dot{x}_{11} + \dot{x}_8 + \sin(\dot{x}_9)) \cdot (-\exp(\dot{x}_0)) $ &  \\
        $0.727$ & $16$ & $-(-\dot{x}_3 + (- \dot{x}_8)) + (\dot{x}_{11} + \dot{x}_5 + \dot{x}_8) \cdot (- \exp(\dot{x}_{11}))   $ &  \\
        $0.749$ & $13$ & $\dot{x}_{11} + \dot{x}_8 + (\dot{x}_{11} + \dot{x}_8 + \dot{x}_9) \cdot (-\exp(\dot{x}_6))            $ &  \\
        $0.886$ & $ 4$ & $-\exp(\sin(\dot{x}_0))                                                                                $ &  \\
        $0.886$ & $ 3$ & $-\exp(\dot{x}_0)                                                                                      $ &  \\
        $0.997$ & $ 2$ & $-\dot{x}_8                                                                                            $ &  \\
        $1.000$ & $ 1$ & $k                                                                                                     $ & $k = 1$ \\
        $1.000$ & $ 1$ & $\dot{x}_4                                                                                             $ &  \\
        \addlinespace
        \bottomrule
    \end{tabularx}
\end{table}

\begin{figure}[htpb]
    \begin{overpic}[width=1.0\textwidth]{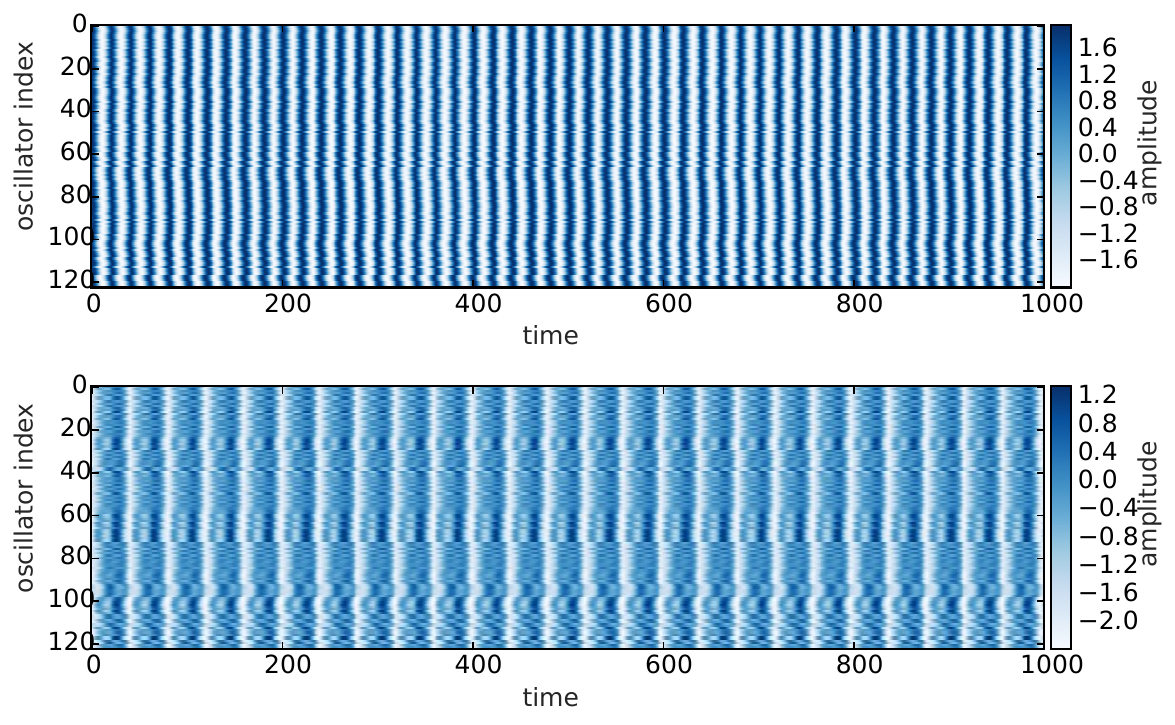}
        \put (0,59.5) {\textbf{(a)}}
        \put (0,29.0) {\textbf{(b)}}
    \end{overpic}
    \caption{Oscillators in a hierarchical network.  \textbf{a:} uncontrolled system; \textbf{b:} Pareto-front solution, $u(\dot{\vec{x}})=\dot{x}_3 + \dot{x}_8 - (\dot{x}_{11} + \dot{x}_8 + \sin(\dot{x}_9)) \cdot \exp(\dot{x}_0)$, for forced de-synchronization.\label{fig:dorogovtsev_vdp_desync_heatmap}}
\end{figure}

\begin{figure}[htpb]
    \includegraphics[width=1.\textwidth]{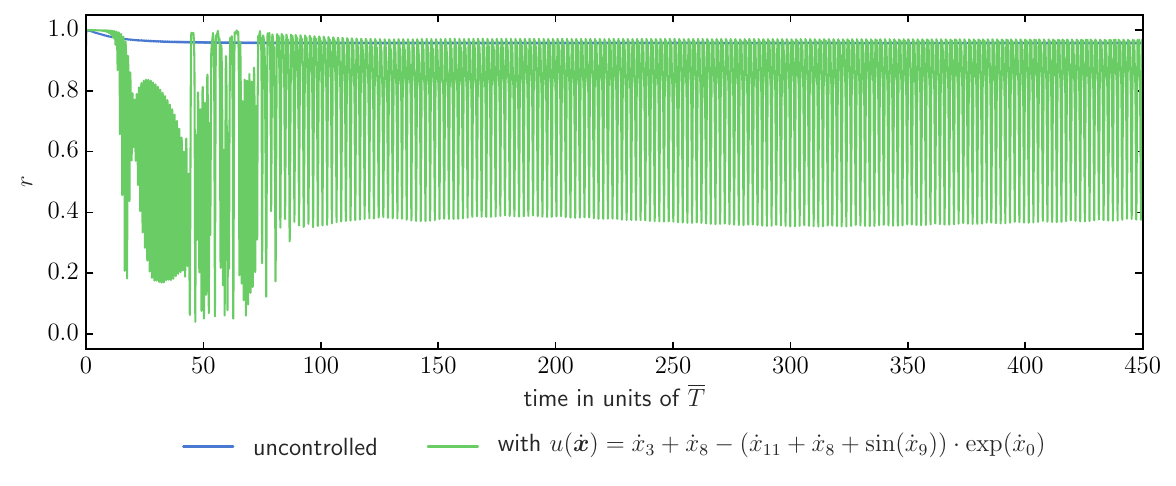}
    \caption{Oscillators in a hierarchical network. Kuramoto order parameter $r$ for the uncontrolled system (blue) and the system controlled by the best solution, with respect to $\Gamma_1$ (green).\label{fig:dorogovtsev_vdp_desync_kuramoto}}
\end{figure}

\section{Conclusions and Future Work}
\label{sec:conc}
Our main question, in this work, concerns the control of a synchronization in systems of coupled oscillators. We presented a computational intelligence-based framework for inferring optimal control laws to achieve this goal. A multi-objective genetic programming algorithm with regression-based constant estimation is used to learn the control laws dynamically.

We first tested our method on a well-known control problem in dynamical systems: Drive a damped harmonic oscillator to a limit cycle and a stable one to a fixed point. We then applied our control approach to dynamical systems composed of networks of coupled oscillators, starting from a system of two coupled van der Pol oscillators up to a hierarchical network consisting of a few hundred oscillators. 
In comparison to other methods like generalized linear regression methods, GP is relatively complex to use for a beginner, however the effort pays off if general solutions are needed. Due to the evolutionary nature of the method, it is not guaranteed that the global optimum is found, which in our case was not a problem, but may well be in other situations.

As a result we find terms of different complexity leading to different levels of synchronization control, where synchronization is measured using the Kuramoto parameter. The results clearly demonstrate the ability of GP-based control to bring a desynchronized system to a synchronization state and vice versa. For forced synchronization, in any setup we find simple control laws, suggesting a single oscillator taking over the control and governing the overall dynamics. For forced desynchronization, laws of increasing complexity are found, where the complexity increased with that of the system complexity.

The results in all setups highlight the importance of designing the objective functions appropriately. In the current setup, we simply relied on learning the control laws by minimizing the error with the desired output. We did not specify any symmetry or energy function to be minimized, nor did we restrict the number of oscillators to be controlled. These details appear to be important for using our methods in a real-world application.
The difficulty in the very general approach is recognized by the asymmetry of the control laws, which is a disadvantage in our opinion. One thus has to carefully analyze the objectives and may upgrade them step by step if unwanted solutions occur.

Further work will extend current methods in several ways: a subsequent automatic stability analysis would be performed for the numerical experiments. This way one can immediately distinguish stable and useful control dynamics from unstable ones. Second, we aim to look into the design of better objective functions, taking into consideration prior domain knowledge, e.g. in the form of additive symmetry terms. Finally, we plan to integrate methods in our framework that would search for the optimal sensor (for measurement) and pressure (for actuation) points in the network for a better control.

\bibliographystyle{unsrt}

\end{document}